\documentstyle[12pt,indent,epsf,eqsection,subeqnarray]{article}

\begin{document}

\bibliographystyle{plain}

\title{Absence of Phase Transition\break
       for Antiferromagnetic Potts Models\break
       via the Dobrushin Uniqueness Theorem
      }
\author{
  {\small Jes\'us Salas}                  \\[-0.2cm]
  {\small Alan D.~Sokal}                  \\[-0.2cm]
  {\small\it Department of Physics}       \\[-0.2cm]
  {\small\it New York University}         \\[-0.2cm]
  {\small\it 4 Washington Place}          \\[-0.2cm]
  {\small\it New York, NY 10003 USA}      \\[-0.2cm]
  {\small\tt SALAS@MAFALDA.PHYSICS.NYU.EDU},
                    {\small\tt SOKAL@NYU.EDU}   \\[-0.2cm]
  {\protect\makebox[5in]{\quad}}  
  \\
}
\vspace{0.5cm}

\maketitle
\thispagestyle{empty}   

\def\spose#1{\hbox to 0pt{#1\hss}}
\def\ltapprox{\mathrel{\spose{\lower 3pt\hbox{$\mathchar"218$}}
 \raise 2.0pt\hbox{$\mathchar"13C$}}}
\def\gtapprox{\mathrel{\spose{\lower 3pt\hbox{$\mathchar"218$}}
 \raise 2.0pt\hbox{$\mathchar"13E$}}}
\def\inapprox{\mathrel{\spose{\lower 3pt\hbox{$\mathchar"218$}}
 \raise 2.0pt\hbox{$\mathchar"232$}}}

\vspace{1cm}
\begin{center}
{\em Dedicated to the memory of R.L.~Dobrushin}
\end{center}

\vspace{1cm}

\begin{abstract}
We prove that the $q$-state Potts antiferromagnet on a lattice
of maximum coordination number $r$ exhibits
exponential decay of correlations uniformly at all temperatures
(including zero temperature) whenever $q > 2r$.
We also prove slightly better bounds for several two-dimensional lattices:
square lattice (exponential decay for $q \ge 7$),
triangular lattice ($q \ge 11$),
hexagonal lattice ($q \ge 4$),
and Kagom\'e lattice ($q \ge 6$).
The proofs are based on the Dobrushin uniqueness theorem.
\end{abstract}


\clearpage

\newcommand{\be}{\begin{equation}}
\newcommand{\ee}{\end{equation}}
\newcommand{\<}{\langle}
\renewcommand{\>}{\rangle}
\newcommand{\para}{\|}
\renewcommand{\perp}{\bot}

\def\smfrac#1#2{{\textstyle\frac{#1}{#2}}}
\def\half{ {{1 \over 2 }}}
\def\smhalf{ {\smfrac{1}{2}} }
\def\scra{{\cal A}}
\def\scrc{{\cal C}}
\def\scrd{{\cal D}}
\def\scre{{\cal E}}
\def\scrf{{\cal F}}
\def\scrh{{\cal H}}
\def\scrk{{\cal K}}
\def\scrm{{\cal M}}
\newcommand{\scrmvec}{\vec{\cal M}_V}
\def\scrmtens{{\stackrel{\leftrightarrow}{\cal M}_T}}
\def\scro{{\cal O}}
\def\scrp{{\cal P}}
\def\scrr{{\cal R}}
\def\scrs{{\cal S}}
\def\ttens{{\stackrel{\leftrightarrow}{T}}}
\def\scrv{{\cal V}}
\def\scrw{{\cal W}}
\def\scry{{\cal Y}}
\def\tauss{\tau_{int,\,\scrm^2}}
\def\taux{\tau_{int,\,{\cal M}^2}}
\newcommand{\taum}{\tau_{int,\,\vec{\cal M}}}
\def\taue{\tau_{int,\,{\cal E}}}
\newcommand{\imag}{\mathop{\rm Im}\nolimits}
\newcommand{\real}{\mathop{\rm Re}\nolimits}
\newcommand{\tr}{\mathop{\rm tr}\nolimits}
\newcommand{\sgn}{\mathop{\rm sgn}\nolimits}
\newcommand{\codim}{\mathop{\rm codim}\nolimits}
\newcommand{\rank}{\mathop{\rm rank}\nolimits}
\newcommand{\sech}{\mathop{\rm sech}\nolimits}
\def\textprime{{${}^\prime$}}
\newcommand{\longto}{\longrightarrow}
\def\var{ \hbox{var} }
\newcommand{\gtilde}{ {\widetilde{G}} }
\newcommand{\USp}{ \hbox{\it USp} }
\newcommand{\CP}{ \hbox{\it CP\/} }
\newcommand{\QP}{ \hbox{\it QP\/} }
\def\hboxscript#1{ {\hbox{\scriptsize\em #1}} }

\newcommand{\plotdot}{\makebox(0,0){$\bullet$}}
\newcommand{\plotsmalldot}{\makebox(0,0){{\footnotesize $\bullet$}}}

\def\bsigma{\mbox{\protect\boldmath $\sigma$}}
\def\bpi{\mbox{\protect\boldmath $\pi$}}
\def\btau{\mbox{\protect\boldmath $\tau$}}
\def\bn{{\bf n}}
\def\br{{\bf r}}
\def\bz{{\bf z}}
\def\bh{\mbox{\protect\boldmath $h$}}

\def\betatilde{ {\widetilde{\beta}} }
\def\hatp{\hat p}
\def\hatl{\hat l}

\def\msbar{ {\overline{\hbox{\scriptsize MS}}} }
\def\normalmsbar{ {\overline{\hbox{\normalsize MS}}} }

\def\eff{ {\hbox{\scriptsize\em eff}} }

\newcommand{\reff}[1]{(\ref{#1})}

\newcommand{\Z}{{\bf Z}}
\newcommand{\zed}{{\bf \Z}}
\newcommand{\R}{\hbox{{\rm I}\kern-.2em\hbox{\rm R}}}
\font\srm=cmr7 		
\def\szed{\hbox{\srm Z\kern-.45em\hbox{\srm Z}}}
\def\sR{\hbox{{\srm I}\kern-.2em\hbox{\srm R}}}
\def\C{{\bf C}}



\newtheorem{theorem}{Theorem}[section]
\newtheorem{corollary}[theorem]{Corollary}
\newtheorem{lemma}[theorem]{Lemma}
\newtheorem{conjecture}[theorem]{Conjecture}
\newtheorem{definition}[theorem]{Definition}
\def\proof{\bigskip\par\noindent{\sc Proof.\ }}
\def\qed{\hbox{\hskip 6pt\vrule width6pt height7pt depth1pt \hskip1pt}\bigskip}

%
%
\newenvironment{sarray}{
          \textfont0=\scriptfont0
          \scriptfont0=\scriptscriptfont0
          \textfont1=\scriptfont1
          \scriptfont1=\scriptscriptfont1
          \textfont2=\scriptfont2
          \scriptfont2=\scriptscriptfont2
          \textfont3=\scriptfont3
          \scriptfont3=\scriptscriptfont3
        \renewcommand{\arraystretch}{0.7}
        \begin{array}{l}}{\end{array}}

\newenvironment{scarray}{
          \textfont0=\scriptfont0
          \scriptfont0=\scriptscriptfont0
          \textfont1=\scriptfont1
          \scriptfont1=\scriptscriptfont1
          \textfont2=\scriptfont2
          \scriptfont2=\scriptscriptfont2
          \textfont3=\scriptfont3
          \scriptfont3=\scriptscriptfont3
        \renewcommand{\arraystretch}{0.7}
        \begin{array}{c}}{\end{array}}


\section{Introduction}  \label{sec_intro}

Dobrushin's uniqueness theorem
\cite{Dobrushin_68,Dobrushin_70,Georgii_88,Simon_93}
provides a simple but powerful method for proving the
uniqueness of the infinite-volume Gibbs measure,
as well as the exponential decay of correlations in this unique Gibbs measure,
for classical-statistical-mechanical systems deep in a single-phase region.
The basic idea underlying this theorem is that
if the probability distribution of a single spin
$\sigma_i$ depends ``sufficiently weakly'' on the remaining spins
$\{\sigma_j\}_{j \neq i}$, then one can deduce
(by a clever iterative argument)
uniqueness of the Gibbs measure and exponential decay of correlations.

The principal applications of this method have been in two regimes:
\begin{itemize}
   \item[1)]  {\em High temperature.}\/  Here $\sigma_i$ depends weakly on
      the $\{\sigma_j\}_{j \neq i}$ because of the strong thermal fluctuations.
   \item[2)]  {\em Large magnetic field.}\/  Here $\sigma_i$ tends to
      follow the magnetic field, no matter what the other spins are doing;
      so the probability distribution of $\sigma_i$
      again depends weakly on the $\{\sigma_j\}_{j \neq i}$.
\end{itemize}
However, Koteck\'y (cited in \cite[pp.~148--149, 457]{Georgii_88}) 
has pointed out that Dobrushin's theorem is applicable also in a third regime:
\begin{itemize}
   \item[3)]  {\em High entropy.}\/  Here $\sigma_i$ has so many states
      available to it (with equal or almost equal probability),
      no matter what the other spins are doing,
      that its probability distribution again depends weakly on the
      $\{\sigma_j\}_{j \neq i}$.
\end{itemize}
The simplest example of this situation is the antiferromagnetic
$q$-state Potts model \cite{Potts_52,Wu_82,Wu_84}
\be 
\scrh   \;=\;   -J \sum_{x \sim y}  \delta_{\sigma_x \sigma_y} 
\label{potts_hamiltonian} 
\ee
with $J=-\beta < 0$,  
on a lattice in which each site has $r$ nearest neighbors.\footnote{
   We use the notation $x \sim y$ to indicate that $x$ is a nearest neighbor 
   of $y$. 
   The sum in \reff{potts_hamiltonian}  
   thus runs over all nearest-neighbor pairs of lattice sites (each 
   pair counted once),
   and each spin takes values $\sigma_x \in \{ 1, 2, \ldots, q \}$.
   The antiferromagnetic case corresponds to $J = -\beta < 0$.
}
Even at zero temperature ($J=-\infty$), 
the spin $\sigma_i$ is required only to be
different from all the neighboring spins $\{\sigma_j\}_{j \sim i}$.
If $q \gg r$, then the probability distribution of $\sigma_i$ depends
only weakly on the values of the $\{\sigma_j\}_{j\sim i}$.
It turns out that Dobrushin's theorem
is applicable whenever $q > 2r$ (see Section \ref{sec3} below),
as well as in some additional cases (see Sections \ref{sec4} and \ref{sec5}).
Thus, for $q$ sufficiently large (how large depends on the
lattice under consideration), the $q$-state Potts antiferromagnet
has a unique Gibbs measure and exponential decay of correlations
at all temperatures, {\em including zero temperature}\/:
the system is disordered as a result of entropy.

\begin{figure}
\epsfxsize=400pt\epsffile{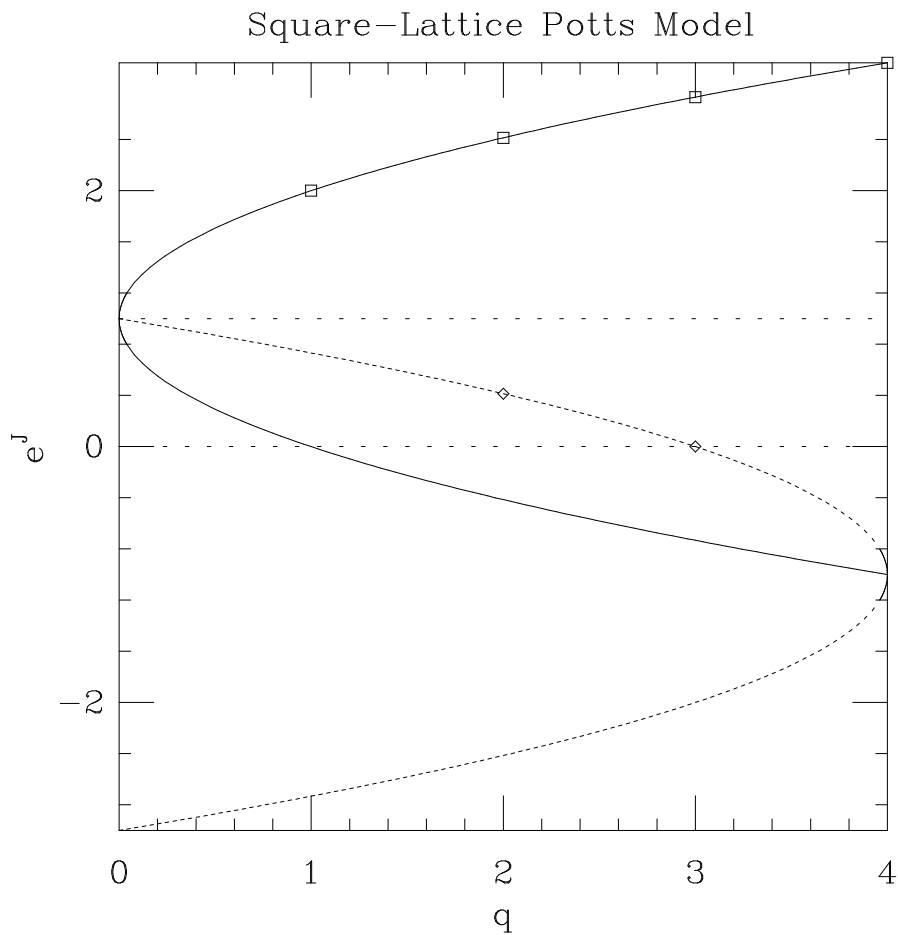}
  \caption[a]{\protect\label{Figure_critical_curves_square}
 Curves where the square-lattice Potts model has been solved: 
 the self-dual curve $(e^J-1)^2=q$ (solid curve), and 
 $(e^J+1)^2 = 4 -q$ (dashed curve).  
 The horizontal dotted lines correspond to $e^J=1$ 
 (separating the ferromagnetic and antiferromagnetic regimes) and to 
 $e^J=0$ (separating the antiferromagnetic regime from the unphysical 
 region $e^J<0$).  
 The squares ($\Box$) show the known ferromagnetic critical points  
 ($q=1,2,3,4$); and the diamonds ($\Diamond$) mark the 
 known antiferromagnetic critical points ($q=2,3$).  
  }
\end{figure}

More precisely, we expect that for each lattice ${\cal L}$ there
will be a value $q_c({\cal L})$ such that
\begin{itemize}
   \item[(a)]  For $q > q_c({\cal L})$  the model has exponential decay
       of correlations uniformly at all temperatures,
       including zero temperature.
   \item[(b)]  For $q = q_c({\cal L})$  the model has a critical point
       at zero temperature.
   \item[(c)]  For $q < q_c({\cal L})$  any behavior is possible.
       Often (though not always) the model has a phase transition
       at nonzero temperature, which may be of either first or second 
       order.\footnote{  
       Exceptions to the usual behavior are, for example, the Ising model 
       ($q=2$) on the triangular lattice ($q_c=4$), which  
       has a zero-temperature critical point \cite{Stephenson_64}; and 
       the Ising model on the Kagom\'e lattice ($q_c=3$), which is non-critical 
       at all temperatures, including zero temperature \cite{Syozi_72}.  
       }
\end{itemize}
Here is what is believed to be true
for the standard two-dimensional lattices:
%
%

{\em Square lattice.}\/
Baxter \cite{Baxter_82,Baxter_82b} has determined the exact free energy
(among other quantities) for the square-lattice Potts model
on two special curves in the $(J,q)$-plane  
(see Figure~\ref{Figure_critical_curves_square}):
\begin{eqnarray}
   e^J   & = &   1 \pm \sqrt{q}             \label{eq1.1}  \\[1mm]
   e^J   & = &  -1 \pm \sqrt{4-q}           \label{eq1.2}
\end{eqnarray}
Curve (\ref{eq1.1}${}_+$) is known to correspond to the ferromagnetic
critical point,
and Baxter \cite{Baxter_82b} has conjectured that
curve (\ref{eq1.2}${}_+$) corresponds to the antiferromagnetic critical point.
For $q=2$ this gives the known exact value \cite{Onsager_44};
for $q=3$ it predicts a zero-temperature critical point ($J_c = -\infty$),
in accordance with previous belief
\cite{Lenard_67,Baxter_70}\footnote{
   Note also that the $q=3$ model is exactly soluble
   at zero temperature in an arbitrary magnetic field
   \cite{Baxter_70,Truong_86,Pearce_89a,Pearce_89b};
   this might increase one's suspicions that the zero-temperature zero-field
   case is critical.
   Indeed, Henley \cite{Henley_94}  
   has some very interesting predictions
   for the critical exponents.
};
and for $q>3$ it predicts that the putative critical point lies in the
unphysical region ($e^{J_c} < 0$), so that the entire physical region
$-\infty \le J \le 0$ lies in the disordered phase.
These predictions for $q=3,4$ have recently been confirmed by
high-precision Monte Carlo simulation \cite{Ferreira-Sokal}.
For some further interesting speculations, see \cite{Saleur_90,Saleur_91}.

\begin{figure}
\epsfxsize=400pt\epsffile{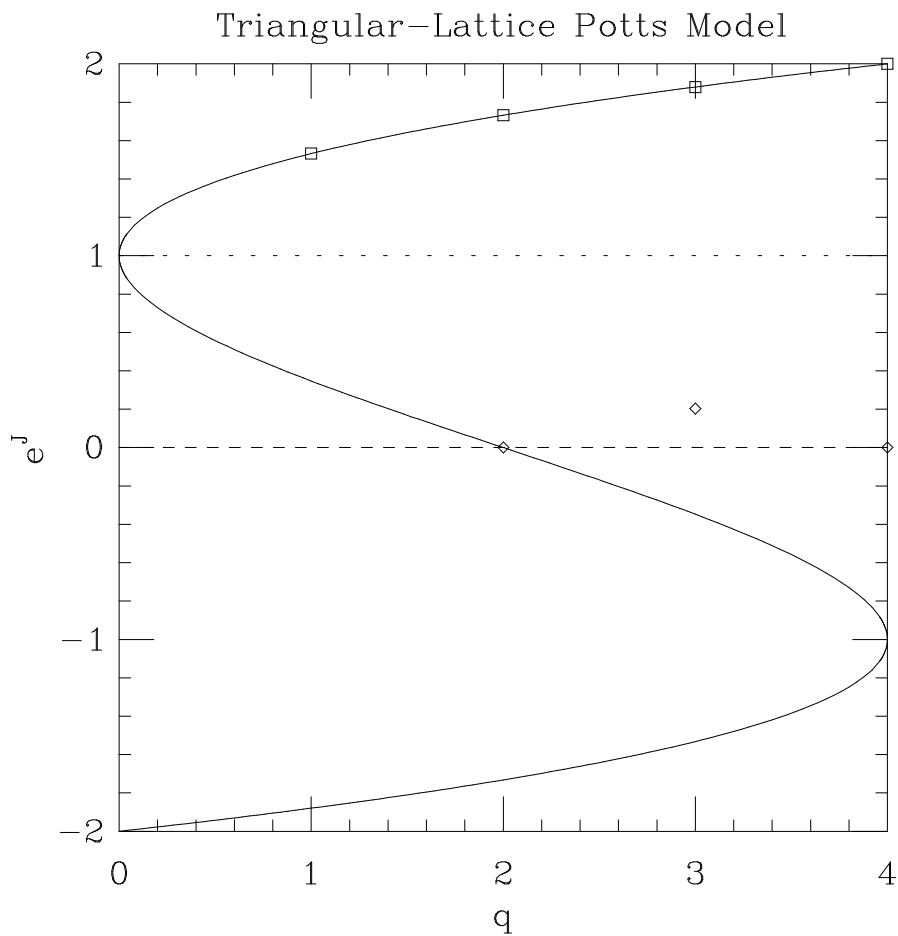}
  \caption[a]{\protect\label{Figure_critical_curves_triangular}
 Curves where the triangular-lattice Potts model has been solved: 
 $(e^J-1)^2(e^J+2)=q$ (solid 
 curve), which has three branches; and the line $e^J=0$ (dashed line). 
 The horizontal dotted line corresponds to $e^J=1$
 (separating the ferromagnetic and antiferromagnetic regimes). 
 The squares ($\Box$) show the known ferromagnetic critical points 
 ($q=1,2,3,4$); and the diamonds ($\Diamond$) mark the
 known antiferromagnetic phase-transition points ($q=2,3,4$).
 Note that the antiferromagnetic first-order transition for $q=3$ does 
 {\em not} lie on either of the exactly-solved curves. 
  }
\end{figure}

{\em Triangular lattice.}\/
Baxter and collaborators \cite{Baxter_78,Baxter_86,Baxter_87}
have determined the exact free energy
(among other quantities) for the triangular-lattice Potts model
on two special curves in the $(J,q)$-plane 
(see Figure~\ref{Figure_critical_curves_triangular}):
\begin{eqnarray}
   (e^J -1)^2 \, (e^J + 2)   & = &   q             \label{eq1.3}  \\[1mm]
   e^J   & = &   0 \qquad\hbox{for } 0 < q < 4     \label{eq1.4}
\end{eqnarray}
The uppermost branch of 
curve \reff{eq1.3} is known to correspond to the ferromagnetic
critical point \cite{Baxter_78},
%
%
and Baxter \cite{Baxter_86} has conjectured that
\reff{eq1.4} corresponds to the antiferromagnetic critical point.
This prediction of a zero-temperature critical point
is known to be correct for $q=2$ \cite{Stephenson_64},
and there is heuristic analytical evidence that it is correct also for $q=4$
\cite{Henley_94,Baxter_70_TRI}.
On the other hand, for $q=3$ this prediction contradicts the rigorous result
\cite{vEFS_unpub}, based on Pirogov-Sinai theory,
that there is a low-temperature phase with long-range order
and small correlation length.
Indeed, a recent Monte Carlo study of the $q=3$ model
has found strong evidence for a first-order transition
(to an ordered phase) at $\beta \approx 1.594$ \cite{Adler_95}.
For $q>4$ one may expect that the triangular-lattice Potts model is noncritical
even at zero temperature.
Finally, the physical meaning of the two lower branches of \reff{eq1.3} is 
mysterious. 
The lowermost branch of \reff{eq1.3} lies entirely in the unphysical  
region 
$e^J<0$. The middle branch is located in the antiferromagnetic region for 
$0<q<2$, and in the unphysical region for $q>2$; at $q=2$ it coincides with 
the antiferromagnetic critical point. For some further interesting speculations,
see \cite{Saleur_90}.

\begin{figure}
\epsfxsize=400pt\epsffile{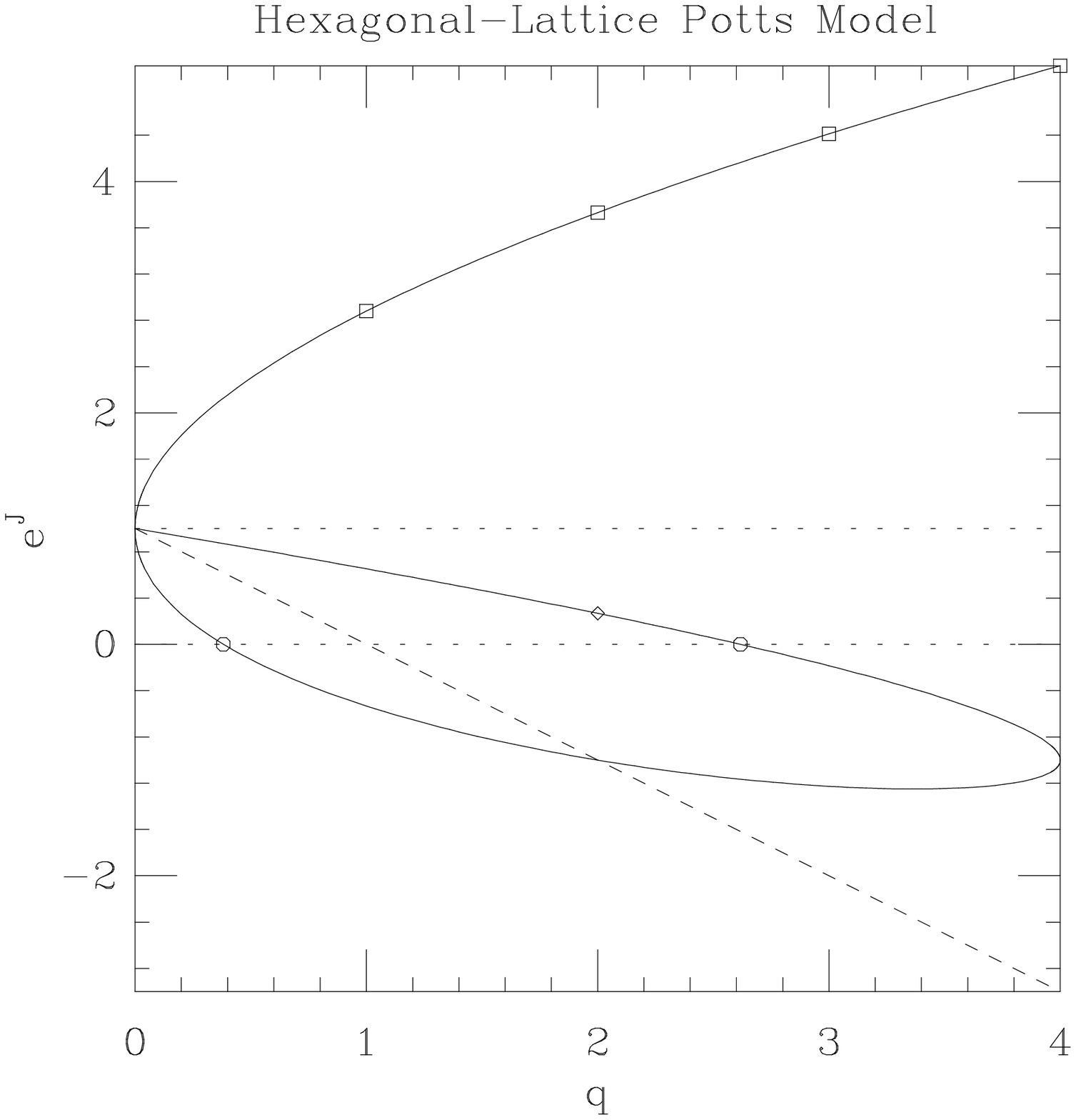}
  \caption[a]{\protect\label{Figure_critical_curves_hexagonal}
 Curves where the hexagonal-lattice Potts model has been solved:
 $(e^J-1)^3 -3q(e^J-1)=q^2$ (solid
 curve), which has three branches; and the line $e^J=1-q$ (dashed line).
 The horizontal dotted lines correspond to $e^J=1$
 (separating the ferromagnetic and antiferromagnetic regimes) and to 
 $e^J=0$ (separating the antiferromagnetic regime from the unphysical region 
 $e^J<0$).
 The squares ($\Box$) show the known ferromagnetic critical points 
 ($q=1,2,3,4$); and the diamond ($\Diamond$) marks the
 known antiferromagnetic critical point for $q=2$.  
 The open circles ($\circ$) show the points where 
 the two antiferromagnetic branches cross the $e^J=0$ line, namely   
 $q = (3 \pm \sqrt{5})/2$.  
  }
\end{figure}

{\em Hexagonal lattice.}\/
This lattice is connected by duality \cite{Baxter_82}
with the triangular lattice\footnote{
   Furthermore, the hexagonal-lattice Potts model
   {\em on the curve \reff{eq1.5}}\/ (and only there)
   can be mapped via the star-triangle transformation
   onto a triangular-lattice Potts model,
   which turns out to lie exactly on the curve \reff{eq1.3}.
};
the image of \reff{eq1.3}/\reff{eq1.4} is
\begin{eqnarray}
   (e^J -1)^3  - 3q(e^J -1) - q^2 & = &   0       \label{eq1.5}  \\[1mm]
   e^J   & = &   1-q  \qquad\hbox{for } 0 < q < 4   \label{eq1.6}
\end{eqnarray}
Curve \reff{eq1.5} has three branches in the region $q \ge 0$ 
(see Figure~\ref{Figure_critical_curves_hexagonal}):
%
%
the uppermost branch (with $0 \le q < \infty$ and $e^J \ge 1$)
is the ferromagnetic critical point;
the middle branch (with $0 \le q \le 4$ and $-1 \le e^J \le 1$)
contains the antiferromagnetic Ising critical point
($q=2$, $e^J = 2 - \sqrt{3}$)
and crosses the zero-temperature point $e^J =0$ at
$q = (3+\sqrt{5})/2 \approx 2.618$;
while the lowermost branch crosses the zero-temperature point $e^J =0$ at
$q = (3-\sqrt{5})/2 \approx 0.382$.\footnote{ 
  The middle branch is missing in \cite[p.~673, Figure~8]{Saleur_90}.  
}
The meaning of this lowermost branch is mysterious,
as is the meaning of \reff{eq1.6}.
But the behavior of the middle branch suggests that it may be
the antiferromagnetic critical curve:
in this case there would be a zero-temperature critical point for
$q = (3+\sqrt{5})/2$  [if this assertion has any meaning\footnote{
   The Potts models for noninteger $q$ can be given a rigorous meaning
   via the mapping onto the Fortuin-Kasteleyn random-cluster model
   \cite{Fortuin-Kasteleyn_69,Fortuin-Kasteleyn_72,Fortuin_72}.
   The trouble is that in the antiferromagnetic case ($J<0$)
   this latter model has negative weights, and so cannot be given
   a standard probabilistic interpretation.
   In particular, the existence of a good infinite-volume limit
   is problematical;  the limit could depend strongly on the
   subsequence of lattice sizes and on the boundary conditions.
   The same is true of the ``anti-Fortuin-Kasteleyn'' representation, 
   in which the coefficients are products of chromatic polynomials of 
   clusters: again the weights can be negative for non-integer $q$, and 
   the existence of the infinite-volume limit is problematical. 
   Likewise, the ice-model representation \cite{Baxter_82,Baxter_76}  
   has in general complex weights for $0<q<4$,  
   even in the ferromagnetic case. 
}],
and the model would be disordered even at zero temperature
for $q > (3+\sqrt{5})/2$. For some further interesting speculations, see 
\cite{Saleur_90}.

{\em Kagom\'e lattice.}\/
This is not merely an academic example, as 
some condensed-matter systems
(for instance, the insulator SrCr$_{8-x}$Ga$_{4-x}$O$_{19}$)
have the Kagom\'e lattice structure \cite{Takano_71,Broholm_91,Huse_92}.
For $q=2$ this model has been solved exactly \cite{Syozi_72},
and there is no phase transition at any temperature.
For $q=3$ the zero-temperature model can be mapped onto the
zero-temperature 4-state triangular-lattice Potts antiferromagnet
\cite{Baxter_70_TRI} 
and so is expected to be critical \cite{Henley_94,Henley_95}.
For $q>3$ one may expect that this model is noncritical
even at zero temperature.

In Table \ref{Table_results} we summarize the believed exact values of
$q_c({\cal L})$ for these four lattices,
along with the upper bounds that follow from our computer-assisted proofs.
Clearly, our rigorous bounds
still fall far short of what is believed to be true in most of the lattices 
considered here. Only for the hexagonal lattice are we somewhat close 
to the expected result.

\bigskip

The plan of this paper is as follows:
In Section \ref{sec2} we set the notation and recall the
Dobrushin uniqueness theorem.
In Section \ref{sec3} we prove that the Dobrushin uniqueness theorem
is applicable to the $q$-state Potts antiferromagnet
on a lattice of maximum coordination number $r$,
uniformly at all temperatures (including zero temperature),
whenever $q > 2r$.
In Section \ref{sec4} we improve this result for some common lattices
(square, hexagonal, triangular, and Kagom\'e),
using a single-site decimation scheme and a computer-assisted proof.
Finally in Section~\ref{sec5} we improve our results for the hexagonal and 
Kagom\'e lattices using more sophisticated decimation schemes.

\bigskip

During the preparation of this paper we learned of the
tragic death of Professor Dobrushin, one of the founders of and
main contributors to modern mathematical statistical mechanics.
We dedicate this paper to his memory.

%
%

\section{Notation and Preliminaries}   \label{sec2}

\subsection{Basic Setup}  \label{sec2.1}

The basic framework for all our results is the
Dobrushin--Lanford--Ruelle (DLR) approach to the
equilibrium statistical mechanics of
infinite-volume classical lattice systems.
A pedagogical introduction to this theory can be found in
\cite[Sections 2.1--2.3]{vEFS_93};
detailed expositions can be found in
the books of Preston \cite{Preston_76}, Georgii  
\cite{Georgii_88} 
and Simon \cite{Simon_93}.
Here we summarize very briefly the notation and the basic ideas.
The central idea in the DLR theory
is to define an {\em infinite-volume Gibbs measure}\/
as a probability distribution for the infinite-volume system
whose {\em conditional}\/ probabilities for {\em finite}\/ subsystems
are given by the Boltzmann-Gibbs formula for the given formal Hamiltonian.

Consider a classical-statistical-mechanical system on a
countably infinite lattice $\cal L$, with spin variables
$\sigma_i$ ($i \in {\cal L}$) taking values in some state space $E$.\footnote{
   In all the applications in this paper,
   the state space $E$ will be the {\em finite}\/ set $\{1,\ldots,q\}$.
   However, the Dobrushin uniqueness theorem is valid in much greater
   generality.
}
The equilibrium statistical mechanics of such a system
is defined by a {\em specification}\/
$\Pi = \{\pi_\Lambda\}_{\Lambda \, \hbox{\scriptsize finite }  
       \subset {\cal L}}$:
here $\pi_\Lambda(\sigma_\Lambda | \sigma_{\Lambda^c})$
gives the conditional probability distribution for the
spin configuration $\sigma_\Lambda \equiv \{\sigma_i\} _{i \in \Lambda}$
{\em inside}\/ the finite set $\Lambda$,
given the spin configuration
$\sigma_{\Lambda^c} \equiv \{\sigma_i\} _{i \in \Lambda^c}$
{\em outside}\/ $\Lambda$.
The $\{\pi_\Lambda\}$ have to satisfy various consistency conditions
\cite{Georgii_88,vEFS_93,Preston_76}.
We shall further assume that each kernel $\pi_\Lambda$
is {\em quasilocal}\/ \cite{Georgii_88,vEFS_93}:
this is a very mild decay condition on the long-range interactions.

Usually the specification $\{\pi_\Lambda\}$ is defined via an
{\em interaction}\/ (= ``formal Hamiltonian'')
$\Phi = \{\Phi_A\}_{A \, \hbox{\scriptsize finite } \subset {\cal L}}$:
here $\Phi_A$ is, roughly speaking, the elementary contribution
to the Hamiltonian coming from the finite set of spins $A \subset {\cal L}$.
Thus, the Hamiltonian $H^\Phi_\Lambda$ for volume $\Lambda$
with external condition $\sigma_{\Lambda^c}$ is
\be
   H^\Phi_\Lambda (\sigma_\Lambda|\sigma_{\Lambda^c})  \;=\;
      \sum_{\begin{scarray}
              A \, \hbox{\scriptsize finite } \subset {\cal L}   \\
              A \cap \Lambda \neq \emptyset
            \end{scarray}
           }
      \Phi_A (\sigma_\Lambda,\sigma_{\Lambda^c}) \; . 
 \label{eq_Phi}
\ee
The kernel $\pi_\Lambda$ is then, by definition,
the corresponding Boltzmann-Gibbs measure:
\be
   \pi^\Phi_\Lambda(\sigma_\Lambda|\sigma_{\Lambda^c})   \;=\;
  Z^\Phi_\Lambda(\sigma_{\Lambda^c})^{-1}
         \exp[-H^\Phi_\Lambda(\sigma_\Lambda|\sigma_{\Lambda^c})] \,
         \prod\limits_{i \in \Lambda} d\mu^0(\sigma_i)
  \;,
 \label{eq_Pi}
\ee
where $\mu^0$ is the {\em a priori}\/ single-spin distribution.
Under mild summability conditions on the interaction $\{\Phi_A\}$,
it can be shown that \reff{eq_Phi}/\reff{eq_Pi}
are well-defined and satisfy all the conditions for a specification,
and furthermore that the $\pi_\Lambda$ are quasilocal \cite{Georgii_88,vEFS_93}.
(In this paper all interactions will be finite-range,
 so the requisite conditions will hold trivially.)

Finally, a probability measure $\mu$ on the configuration space
of the infinite-volume system is said to be an
{\em infinite-volume Gibbs measure}\/ for the specification $\Pi$
if, for each finite subset $\Lambda \subset {\cal L}$,
the conditional probability distribution
$\mu( \,\cdot\, | \sigma_{\Lambda^c})$
equals $\pi_\Lambda( \,\cdot\, | \sigma_{\Lambda^c})$.
See \cite{Georgii_88,vEFS_93,Preston_76} for details.

\subsection{Dobrushin's Uniqueness Theorem}   \label{sec2.2}

Let us now focus on the kernels $\pi_{\{i\}}$ ($i \in {\cal L}$),
which give the probability distribution of a {\em single}\/ spin $\sigma_i$
conditional on the remaining spins $\{ \sigma_k \} _{k \neq i}$.
Let us begin by fixing a site $i \in {\cal L}$
and another site $j \neq i$.
We shall define a quantity $c_{ij}$ that measures the
strength of direct dependence of $\sigma_i$ on $\sigma_j$:
\be
   c_{ij}   \;\equiv\;
   \sup_{ \{\sigma\}, \{\widetilde{\sigma}\} \colon\;  
                 \sigma_k = \widetilde{\sigma}_k \, \forall k \neq j} 
   d\left(\,  \pi_{\{i\}}(\; \cdot \; | \{\sigma\}) \, , \; 
              \pi_{\{i\}}(\; \cdot \; | \{\widetilde{\sigma}\}) \, \right)  
   \;,
 \label{def_cij}
\ee
where
\be
   d(\mu_1,\mu_2)   \;\equiv\;  \sup_{A \subset E}  |\mu_1(A) - \mu_2(A)|
      \;=\;  \sup_{A \subset E}  [\mu_1(A) - \mu_2(A)]
\label{var_distance} 
\ee
is half the variation distance between the probability measures
$\mu_1$ and $\mu_2$,
and the supremum in \reff{def_cij} is taken over all pairs of configurations
$\{ \sigma_k \} _{k \neq i}$ and $\{ \widetilde{\sigma}_k \} _{k \neq i}$
that differ only at the site $j$.
The matrix $C = (c_{ij})_{i,j \in {\cal L}}$
is called {\em Dobrushin's interdependence matrix}\/.
Please note that $c_{ij}$ is a ``worst-case'' measure of the
dependence of $\sigma_i$ on $\sigma_j$,
in the sense that it is defined via the {\em supremum}\/
over all configurations of the spins
$\{ \sigma_k = \widetilde{\sigma}_k \} _{k \neq i}$,
$\sigma_j$ and $\widetilde{\sigma}_j$.
Finally, we define the {\em Dobrushin constant}\/
\be
   \alpha   \;\equiv\;   \sup_{i \in {\cal L}}  \sum_{j \neq i}  c_{ij}
   \;.
 \label{def0_alpha}
\ee
We then have the following result \cite{Dobrushin_68}:

\begin{theorem}[Dobrushin uniqueness theorem]   \label{thm2.1}
Let $\Pi$ be a {\em quasilocal}\/ specification
whose Dobrushin constant $\alpha$ is $< 1$.
Then there is at most one infinite-volume Gibbs measure for $\Pi$.
\end{theorem}

\noindent
For a proof, see \cite[Section 8.1]{Georgii_88}
or \cite[Section V.1]{Simon_93}.\footnote{
   {\em Warning:}\/  Simon \cite{Simon_93} denotes by $\rho_{ji}$
   what we have called $c_{ij}$ --- note the reversal of indices!
}

{\bf Remarks.}
1. Under very mild conditions on the specification $\Pi$
--- which always hold if, for example, the state space $E$ is finite ---
it can be shown that there exists {\em at least one}\/
infinite-volume Gibbs measure for $\Pi$.
So the upshot of Dobrushin's uniqueness theorem
is that there exists {\em exactly one}\/
infinite-volume Gibbs measure for $\Pi$.

2. There is an extension of Dobrushin's uniqueness theorem that uses the 
Kantorovich--Rubinstein--Vasershtein--Ornstein distance corresponding to an 
arbitrary metric on the state space $E$, in place of the variation 
distance \reff{var_distance}: see \cite{Dobrushin_70} or 
\cite[Section~V.3]{Simon_93}. This extension is particularly useful in 
studying continuous-spin systems. But in the Potts case it gains nothing, 
as the color-permutation symmetry of the Potts Hamiltonian ensures that the 
variation distance is in fact the ``natural'' distance.

\bigskip

The hypotheses of Theorem \ref{thm2.1}
imply also a strong result on the decay of correlations
in the unique infinite-volume Gibbs measure.
We need a few definitions:
For any function $f(\{\sigma\})$ and any site $i$,
we define the {\em oscillation of $f$ at $i$}\/:
\be
  \delta_i(f)   \;\equiv\;
   \sup_{ \{\sigma\}, \{\sigma'\} \colon\;  
                 \sigma_k = \sigma'_k \, \forall k \neq i} 
   \bigl| f(\{\sigma\}) - f(\{\sigma'\}) \bigr|
  \;.
\ee
We say that $f$ has {\em finite total oscillation}\/ if
\be
   \Delta(f)   \;\equiv\;  \sum_{i \in {\cal L}}  \delta_i(f)   \;<\; \infty
   \;.
\ee
(In particular, any bounded function depending on only finitely many spins
has finite total oscillation.)
Finally, let $C^n$ be the $n^{th}$ matrix power of
Dobrushin's interdependence matrix $C$,
and define
\be
   D_{ij}   \;\equiv\;  \sum_{n=0}^\infty  (C^n)_{ij}
   \;.
\ee
We then have:

\begin{theorem}   \label{thm2.2}
Let $\Pi$ be a quasilocal specification
satisfying the Dobrushin condition $\alpha < 1$.
Then the unique infinite-volume Gibbs measure $\mu$ satisfies
\be
   |\mu(fg) - \mu(f) \mu(g)|
   \;\le\;
   {1 \over 4}  \sum_{i,j \in {\cal L}}  \delta_i(f) \, D_{ij} \, \delta_j(g)
\ee
for all functions $f,g$ of finite total oscillation.
\end{theorem}

\noindent
In particular, if $\cal L$ is a regular lattice and the
interaction is of finite range
(so that $c_{ij} = 0$ whenever $|i-j| > R$),
then Dobrushin's condition $\alpha < 1$
implies that $D_{ij}$ decays {\em exponentially}\/ as $|i-j| \to \infty$,
so that Theorem \ref{thm2.2} implies the
{\em exponential decay of correlations}\/
in the unique infinite-volume Gibbs measure.
For proofs of Theorem \ref{thm2.2} as well as these related results,
see \cite[Section 8.2]{Georgii_88} or \cite[Section V.2]{Simon_93}.

%
%

\section{General Proof of Uniqueness for $q>2r$}   \label{sec3}

Let us now apply Dobrushin's uniqueness theorem to a $q$-state
Potts antiferromagnet defined by the (formal) Hamiltonian
\be 
   \scrh   \;=\;   - \sum_{i \sim j}  J_{ij}  \delta_{\sigma_i \sigma_j} 
 \label{eq3.1}
\ee
with all couplings satisfying $-\infty \le J_{ij} \le 0$.
We say that $j$ is a nearest neighbor of $i$
(denoted $j \sim i$) in case $J_{ij} \neq 0$.
We need to calculate the Dobrushin interdependence constants $c_{ij}$.

First, some preliminaries:
Let $\rho$ be a probability measure on the state space $E$,
and let $f \ge 0$ be any function on $E$ such that
$\rho(f) \equiv \int\! f \, d\rho > 0$.
Then we define the probability measure $\rho^{(f)}$ by
\be
   \rho^{(f)}   \;=\;   {f\rho  \over \rho(f)}
\ee
(``$\rho$ weighted by $f$ and then normalized'').

\begin{lemma}  \label{lemma3.1}
Let $0 \le f,g \le 1$.  Then
\be
   d( \rho^{(f)}, \rho^{(g)} )
   \;\le\;
   \max\!\left[  {\rho(1-f) \over \rho(f)} \,,\; 
                 {\rho(1-g) \over \rho(g)}
         \right]   \;.
\ee
\end{lemma}

\proof
By definition,
\be
   d( \rho^{(f)}, \rho^{(g)} )
   \;=\;
   \sup_{A \subset E}  \int_A
   \left[  {g(x) \over \rho(g)}  \,-\,  {f(x) \over \rho(f)}  \right]
   \, d\rho(x)  \;.
\ee
Suppose (without loss of generality) that $\rho(f) \le \rho(g)$.
Then $1/\rho(f) \ge 1/\rho(g)$, so that
\begin{eqnarray}
   {g(x) \over \rho(g)}  \,-\,  {f(x) \over \rho(f)}
   & \le &
   {1-f(x) \over \rho(f)}  \,-\,  {1-g(x) \over \rho(g)}   \\[1mm]
   & \le &
   {1-f(x) \over \rho(f)}  \qquad[\hbox{since }  g \le 1]  \;.
\end{eqnarray} 
But since $f \le 1$, we have
$\int\limits_{A} [1-f(x)] \, d\rho(x)  \le \int\limits_{E} [1-f(x)] \, d\rho(x)
 \equiv \rho(1-f)$.
\qed

{\bf Remark.}
1. If $\rho(f) = \rho(g)$ and the supports of $1-f$ and $1-g$ are disjoint,
then this estimate is sharp.

\bigskip

Let us now apply this lemma to compute the Dobrushin constant $c_{ij}$
in the Potts antiferromagnet \reff{eq3.1}.
We shall assume that the site $i$ has at most $r$ nearest neighbors.
Fix two configurations
$\{ \sigma_k \} _{k \neq i}$ and $\{ \widetilde{\sigma}_k \} _{k \neq i}$
that differ only at the site $j$.
Let $\rho$ be the conditional probability distribution at site $i$
in the presence of all of $i$'s neighbors {\em other than $j$}\/:
\be
   \rho(\sigma_i)  \;=\;
   { \exp\!\left[\sum\limits_{k \neq i,j} J_{ik} \delta_{\sigma_i \sigma_k}
           \right]
     \over
     \sum\limits_{\sigma=1}^q
     \exp\!\left[\sum\limits_{k \neq i,j} J_{ik} \delta_{\sigma \sigma_k}
           \right]
   }
   \;.
\ee
And let
\begin{subeqnarray}
    f(\sigma_i)  \;=\;  \exp[J_{ij} \delta_{\sigma_i \sigma_j}]    \\[2mm]
    g(\sigma_i)  \;=\;  \exp[J_{ij} \delta_{\sigma_i \widetilde{\sigma}_j}]
\end{subeqnarray}
The antiferromagneticity condition $J_{ij} \le 0$ guarantees that
$0 \le f,g \le 1$, so we can apply Lemma \ref{lemma3.1}.
Because the site $i$ has at most $r-1$ nearest neighbors $k \neq j$,
it follows that in the measure $\rho$ there are at least $q-r+1$ states with
equal weight
(namely, those states not equal to any of the $\{ \sigma_k \} _{k \neq i,j}$).
Moreover, since all the $J_{ik}$ are $\le 0$,
all the states that {\em are}\/ equal to one or more of the
$\{ \sigma_k \} _{k \neq i,j}$ have {\em smaller}\/ weight.
Hence the maximum weight given by $\rho$ to any single state is
$\le 1/(q-r+1)$.  Furthermore, $1-f$ is nonzero on at most one state
(namely, $\sigma_j$) and is there $\le 1$;
so we have $\rho(1-f) \le 1/(q-r+1)$.  The same holds for $\rho(1-g)$.
Hence, by Lemma \ref{lemma3.1},
\be
   d( \rho^{(f)}, \rho^{(g)} )
   \;\le\;
   \max\!\left[  {\rho(1-f) \over \rho(f)} \,,\;
                 {\rho(1-g) \over \rho(g)}
         \right]
   \;\le\;
   {1 \over q-r}
   \;,
\ee
so that
\be
   c_{ij}  \;\le\;  {1 \over q-r}
   \label{bound_cij}
\ee
and consequently
\be
   \sum_{j \sim i}  c_{ij}  \;\le\;  {r \over q-r}
   \;.
\ee

If we now assume that {\em every}\/ site in the lattice has
at most $r$ nearest neighbors, we can conclude that
\be
   \alpha   \;\equiv\;   \sup_{i \in {\cal L}}  \sum_{j \sim i}  c_{ij}
            \;\le\;  {r \over q-r}
   \;.
 \label{bound_alpha}
\ee
In particular, Dobrushin's condition $\alpha < 1$ holds whenever $q > 2r$.
Moreover, the bound \reff{bound_alpha} holds {\em uniformly}\/
in the values of the couplings $\{ J_{ij} \}$,
provided only that they are antiferromagnetic.
Therefore, Theorems \ref{thm2.1} and \ref{thm2.2} guarantee that
for $q > 2r$ there exists a unique infinite-volume Gibbs measure
at all temperatures (including zero temperature),
and that this unique infinite-volume Gibbs measure
exhibits exponential decay of correlations uniformly in the temperature.

{\bf Remarks.}
1. For the Potts antiferromagnet at {\em zero}\/ temperature,
i.e.\ when all nonzero $J_{ij}$ equal $-\infty$,
the bound \reff{bound_cij} is sharp.

2. Koteck\'y (cited in \cite[pp.~148--149, 457]{Georgii_88}) obtained the  
result \reff{bound_cij}--\reff{bound_alpha} {\em at zero temperature}. 
But at nonzero temperature he obtained the weaker result 
$c_{ij} \leq 2/(q-r)$ and hence $\alpha\leq 2r/(q-r)$, so that he proved 
uniqueness only for $q>3r$.

%
%
\section{Improvements via Single--Site Decimation}   
\label{sec4}

In this section we are going to improve on the bound $q > 2r$, using a 
computer-assisted proof that must be carried separately for each lattice. 
For each of the four lattices we study (square, hexagonal, triangular, and 
Kagom\'e) we find that uniqueness holds for $q > 2r-2$. But of course there 
is no guarantee that this result holds for general lattices!  

We emphasize that our proof in this section is valid only at zero 
temperature ($J=-\infty$). Presumably the result holds also for finite 
$-\infty <J<0$, but we do not have any proof of this fact.

The idea of our proof is simple: decimate the original lattice and then 
apply Dobrushin's criterion to the decimated lattice. This trick has 
also been used by other authors in a different context \cite{Kennedy_95}.

The decimation step can be expressed in a general fashion. 
Consider a spin $s$ that interacts with $r$ nearest-neighbor spins  
$t_1,\ldots,t_r$. (See Figure~\ref{figure_decimation_square} for the case of 
the square lattice, which has $r=4$.)     
We have to perform the sum $\sum_{s=1}^q \exp (J \sum_j \delta_{s,t_j})$; 
this will give us the statistical weight associated 
after decimation to the spin configuration 
$(t_1,\ldots,t_r)$.  
The result is very simple for the antiferromagnetic model at zero temperature 
(that is, $J=-\infty$): 
\be
\sum_{s=1}^q \exp \! \left( J \sum_{j=1}^r \delta_{s,t_j} \right)    
\; \stackrel{J=-\infty}{=} \;  
\sum_{s=1}^q \prod_{j=1}^r (1 - \delta_{s,t_j}) 
\; =  \; q - C(t_1,\ldots,t_r) \; , 
\label{decimation_factor} 
\ee 
where $C(t_1,\ldots,t_r)$ is the number of distinct spin values (``colors'') 
we have in the configuration $(t_1,\ldots,t_r)$.  
Thus, decimating the spin $s$ will generate the $r$-body interaction 
\reff{decimation_factor} among the spins $t_1,\ldots,t_r$.

In this section we will be considering only regular lattices, in which each 
site has the same number $r$ of nearest neighbors.

\subsection{Square lattice}

In this case the original lattice 
is bipartite, so we can sum over all the spins belonging to one of the two 
sublattices (for instance, the empty circles in  
Figure~\ref{figure_square_lattice}). 
In this way we obtain a 
decimated lattice defined by the rest of the original spins (solid circles 
in Figure~\ref{figure_square_lattice}). 
This decimated lattice is again a square lattice, but 
rotated 45 degrees with respect to the original one.

The previous discussion on the decimation procedure  
tells us that the interaction on the decimated lattice 
lives on the ``plaquettes'' (= squares): the statistical weight for each  
such square is given by \reff{decimation_factor}.  
Each spin $t_0$ on the decimated lattice  
(see Figure~\ref{figure_square_lattice}) interacts with the other eight  
spins located on the four squares to which $t_0$ belongs:   
$(t_0,t_1,t_2^\prime,t_2)$,     $(t_0,t_2,t_3^\prime,t_3)$,  
$(t_0,t_3,t_4^\prime,t_4)$, and $(t_0,t_4,t_1^\prime,t_1)$.  
These eight spins fall into two classes: four nearest-neighbor spins  
$t_1,t_2,t_3,t_4$ (which belong simultaneously to two of those squares),  
and four next-to-nearest neighbors 
$t_1^\prime,t_2^\prime,t_3^\prime,t_4^\prime$  
(which belong to only one of those squares). 
Thus, the quantity  
\be 
c_{0,j} \; \equiv \;  \sup_{ \{t\},\{\widetilde{t}\} \colon\;  
                 t_k = \widetilde{t}_k \, \forall k \neq j} 
          d\left(\;  \rho_0(\; \cdot \; | \{t\}) \; , \; 
                \rho_0(\; \cdot \; | \{\widetilde{t}\}) \; \right)  
\label{def_coj} 
\ee
will depend on whether the spin $t_j$
is a nearest neighbor or a next-to-nearest neighbor of the spin $t_0$. 
In what follows, $c_{0,{\rm nn}}$ will denote this quantity evaluated at a 
nearest-neighbor spin $t_j$,  
and $c_{0,{\rm nnn}}$ will denote the same quantity 
evaluated at a next-to-nearest-neighbor spin $t_j$.

To obtain $c_{0,j}$ we have to consider all the $q^8$ distinct configurations 
$\{ t \}$ of the spins   
$t_1,t_2,t_3,t_4,t_1^\prime,t_2^\prime,t_3^\prime,t_4^\prime$,   
compute the conditional probability measure $\rho_0(\, \cdot\, | \{t\})$  
for each such configuration $\{ t \}$,  
compute the variation distance between all pairs of such measures whose 
second arguments differ only by the value of the spin $t_j$,  
and finally take the maximum of those distances.   
In the original proof of Koteck\'y one could easily figure out  
which were the configurations which maximize \reff{def_coj}. Here, this 
is more complicated due to the form of the 4-body interaction 
\reff{decimation_factor}.    
The important point is that for each fixed $q$ there is only a {\em finite} 
number of configurations to look at.    
So we can write a computer algorithm to examine all the possible 
configurations and compute \reff{def_coj}.  
We have written a {\sc Fortran} code implementing these ideas.  
In this case, the number of configurations is manageable, but 
in order to streamline the computation we have exploited the color-permutation 
symmetry of the Potts Hamiltonian and have considered 
only those configurations that are not related by a 
mere relabeling of the colors. 
This list of configurations was generated by another 
{\sc Fortran} code using a recursive algorithm.\footnote{ 
  Given a list of all the possible configurations (not related by a permutation 
  of the colors) for $n$ spins, it  
  is very simple to construct the same list for $n+1$ points. The  
  starting point of the algorithm is trivial: for one spin there is only 
  one such configuration. 
}

For each $q$ we obtained $c_{0,{\rm nn}}$ and $c_{0,{\rm nnn}}$. 
Given these values it is easy to compute the quantity 
\be 
\alpha \equiv \sum_{j\neq0} c_{0,j} = 4c_{0,{\rm nn}} + 4c_{0,{\rm nnn}} \; . 
\label{def_alpha}
\ee
When $\alpha<1$, Dobrushin's theorem states that the 
infinite-volume Gibbs measure is unique and that this measure exhibits  
exponentially decaying correlations. 
We performed this computation for $q=5,6,7,8$. 
[The case $q=4$ is very special. 
First, the statistical weight \reff{decimation_factor} associated to a 
plaquette with all the spins in different colors (i.e. $C=4=q=r$) is zero.  
Second and more important, there are configurations of the spins 
$t_1,\ldots,t_4,t_1^\prime,\ldots,t_4^\prime$ for which {\em all} possible 
values of $t_0$ are forbidden at $T=0$, so that the probability measure 
$\rho_0(\, \cdot \, | \{t\})$ at $T=0$ is ill-defined.  
In this case, 
we would have to compute $\rho_{0}(\, \cdot \,|\{t\})$ at $T>0$   
and then take the limit 
$T\rightarrow0$. We are not going to consider such pathological cases in this 
paper.\footnote{
   In any case, 
   we shall see (empirically for our lattices) that Dobrushin's criterion 
   is never satisfied 
   when $q=r+1$. As $\alpha$ seems (again empirically) to be  
   a decreasing function of $q$, 
   Dobrushin's criterion would not hold when $q=r$.  
}]

The numerical results for $q=5,6,7,8$ 
are displayed in Table \ref{Table_results_square}.  
Moreover, the general formulae for $q\geq6$ can be easily guessed.  
The method is as follows: 
First, we identify which are the configurations that maximize
$c_{0,{\rm nn}}$ for each value of $q$.
There is (empirically) a value of $q=q_{\rm min}^{\rm nn}$ 
such that whenever $q\geq q_{\rm min}^{\rm nn}$   
we always find the same maximizing configurations for $c_{0,{\rm nn}}$.  
For these configurations we can compute 
exactly the value of $c_{0,{\rm nn}}$ for general $q\geq q_{\rm min}^{\rm nn}$. 
The same procedure can be carried out for $c_{0,{\rm nnn}}$. For the 
square lattice we find that $q_{\rm min}^{\rm nn}=6$  and 
$q_{\rm min}^{\rm nnn}=5$. The configurations found to maximize 
$c_{0,{\rm nn}}$ and $c_{0,{\rm nnn}}$ are depicted in    
Table~\ref{Table_configurations_square}. 
{}From these patterns it is very easy to compute the general formulae:
\begin{eqnarray}
c_{0,{\rm nn}} &=& { (q-3)^2 (2q -7) \over
             q^5 - 16q^4 + 108 q^3 - 391 q^2 + 764 q - 639 } \; 
 \quad \hbox{for } q \geq 6 \\[2mm] 
c_{0,{\rm nnn}} &=& { (q-3)^3 \over
             q^5 - 16q^4 + 108 q^3 - 389 q^2 + 749 q - 611 } \; 
 \quad \hbox{for } q \geq 5 \;  
\end{eqnarray}

These results show that Dobrushin's condition 
$\alpha<1$ holds for $q\geq7$. This 
value is two units smaller than the value obtained by Koteck\'y ($q\geq9$), 
although still far from the truth ($q\geq 4$, or more precisely $q>3$).

\subsection{Hexagonal lattice}

This lattice is also bipartite, so we can again sum over one of 
the two sublattices 
(empty circles in Figure~\ref{figure_hexagonal_lattice}). 
By this decimation process we obtain a triangular lattice 
(solid circles in Figure~\ref{figure_hexagonal_lattice}). 
The statistical weight \reff{decimation_factor} consists of 3-body 
interactions living on the triangles that contain a decimated spin in 
their interior. 
To each spin $t_0$ there correspond 
three such triangles: $(t_0,t_1,t_2)$, $(t_0,t_3,t_4)$, and $(t_0,t_5,t_6)$;  
so $t_0$ interacts with six nearest-neighbor spins.  All these spins are 
equivalent, so in this case we only have to compute one quantity 
$c_{0,{\rm nn}}$. We then have $\alpha = 6 c_{0,{\rm nn}}$.

The numerical results for $q=4,5,6$ 
are contained in Table~\ref{Table_results_hexagonal}.  
The general form of $c_{0,{\rm nn}}$ can be guessed from the 
configuration which minimizes $c_{0,{\rm nn}}$ for $q\geq4$; 
this configuration is shown in Table~\ref{Table_configurations_hexagon}.  
The formula for $c_{0,{\rm nn}}$ is
\be
c_{0,{\rm nn}} = { (q-2)^2 \over
                 q^4 - 9 q^3 + 33 q^2 - 59 q +43 }  \quad \hbox{for } 
                q\geq 4 \; .
\ee

We see that Dobrushin's condition $\alpha<1$ holds for $q\geq5$. 
We again improve Koteck\'y's result ($q\geq7$) by two units. 
This should be compared to the believed exact result $q\geq3$ (more precisely, 
$q>2.618\ldots$).

\subsection{Triangular lattice}

This lattice is tripartite. We can decimate it by summing over all the  
spins belonging to one of the three sublattices 
(empty circles in Figure~\ref{figure_triangular_lattice}).  
The result of the decimation process is a  
hexagonal lattice (solid circles in 
Figure~\ref{figure_triangular_lattice}).  
{}From \reff{decimation_factor} we see that the interaction 
lives now on the hexagonal faces of this lattice, so each spin 
$t_0$ interacts with the other 12 spins belonging to the three hexagons 
to which $t_0$ belongs: $(t_0,t_1,t_1^\prime,t_2^\prime,t_3^\prime,t_2)$, 
$(t_0,t_2,t_4^\prime,t_5^\prime,t_6^\prime,t_3)$, and 
$(t_0,t_3,t_7^\prime,t_8^\prime,t_9^\prime,t_1)$.  
There are two types of neighboring spins: 
nearest-neighbor spins $t_1,t_2,t_3$ 
(which belong to two different hexagons), and next-to-nearest neighbors 
$t_1^\prime,\ldots,t_9^\prime$ (which belong to only one hexagon).\footnote{ 
   {\em Geometrically}\/  there are two classes of  
   next-to-nearest-neighbor spins:
   those diametrically opposite to $t_0$ (e.g. $t_2^\prime$) and those not 
   (e.g. $t_1^\prime$ and $t_3^\prime$). But these two classes play identical 
   roles in the interaction \reff{decimation_factor}, which is 
   invariant under permutations of $t_1,\ldots,t_r$. 
}  
For each type we have to compute the corresponding quantity \reff{def_coj}.  
We again denote these $c_{0,{\rm nn}}$ and $c_{0,{\rm nnn}}$, respectively.  
The quantity $\alpha$ can be written as
$\alpha = 3 c_{0,{\rm nn}} + 9 c_{0,{\rm nnn}}$.

There is one important point concerning this lattice. In the previous two 
examples the links of the decimated lattice did not coincide with those of 
the original lattice. 
However, in the triangular lattice the links of the 
decimated lattice are a {\em subset} of the links of the  
original lattice. This means 
that the statistical weight associated to a given hexagon is not 
given merely by \reff{decimation_factor}; one has also the 2-body 
interactions from the original Hamiltonian.  
For example, the weight associated at $T=0$ to
the hexagon $(t_0,t_1,t_1^\prime,t_2^\prime,t_3^\prime,t_2)$ is not 
$q - C(t_0,t_1,t_1^\prime,t_2^\prime,t_3^\prime,t_2)$, but rather  
\be 
 [q - C(t_0,t_1,t_1^\prime,t_2^\prime,t_3^\prime,t_2)] (1-\delta_{t_0,t_1}) 
 (1-\delta_{t_1,t_1^\prime}) (1-\delta_{t_1^\prime,t_2^\prime}) 
 (1-\delta_{t_2^\prime,t_3^\prime}) (1-\delta_{t_3^\prime,t_2})       
 (1-\delta_{t_2,t_0}) \; .  
\ee
When we take account of the three hexagons adjoining $t_0$, we have to  
include 15 factors $1-\delta_{t_k,t_j}$ in our 
statistical weight. 
However, only the three factors 
$(1-\delta_{t_0,t_1})(1-\delta_{t_0,t_2})(1-\delta_{t_0,t_3})$  
are essential. This is because those delta functions 
whose arguments are both distinct from $t_0$ are simply boundary conditions  
(their values are independent of $t_0$). If their product is non-zero,  
they will factor out when computing $\rho_{0}$. If their product is zero, 
then we should go to $T>0$ [where the corresponding weight 
$\exp(J\delta_{t_k,t_j})$ is nonzero], do the 
computation (and factor their contribution out), and take the limit 
$T\rightarrow0$. At the end, the result would be the same as if we had 
omitted these factors from the very beginning.

The numerical results for $7\leq q \leq 12$ are  
displayed in Table~\ref{Table_results_triangular}. 
We find that there are two different types of configurations maximizing
$c_{0,{\rm nn}}$, and only one for $c_{0,{\rm nnn}}$; these patterns
are depicted
in Table~\ref{Table_configurations_triangular}.
Using the configurations represented in
Table~\ref{Table_configurations_triangular}, it is very easy to guess
the general formulae for $c_{0,{\rm nn}}$ and $c_{0,{\rm nnn}}$:
\begin{eqnarray}
c_{0,{\rm nn}} &=& { (q-5)^3 \over
             q^4 - 21 q^3 + 168 q^2 - 609 q + 847 } \; 
 \quad \hbox{for } q \geq 7 \\[2mm] 
c_{0,{\rm nnn}} &=& { (q-5)^3 \over
             q^4 - 21 q^3 + 168 q^2 - 608 q + 841 } \; 
 \quad \hbox{for } q \geq 7 \;  
\end{eqnarray}

We see that Dobrushin's condition $\alpha<1$ holds for 
$q \geq 11$, which is again an improvement of two units compared to 
Koteck\'y's result ($q\geq13$). 
Our result should be compared with the expected exact  
value $q\geq5$ (or more precisely $q>4$).

\subsection{Kagom\'e lattice}

In this case we sum over those spins situated on the top vertex of 
the up-pointing triangles (open circles in 
Figure~\ref{figure_kagome_lattice}). 
After decimation we obtain a square lattice 
defined by the solid circles in Figure~\ref{figure_kagome_lattice}. 
Obviously the interaction 
\reff{decimation_factor} lives on the ``crossed'' squares (i.e.\  those 
which have a decimated spin, indicated by an open circle, inside). 
Each spin $t_0$ interacts with two such 
squares: $(t_0,t_1,t_1^\prime,t_2^\prime)$ and 
$(t_0,t_2,t_3^\prime,t_4^\prime)$. Among the six spins with which 
$t_0$ interacts, 
we can distinguish two types: two nearest-neighbor spins 
($t_1$ and $t_2$), which 
are connected to $t_0$ through an original link; 
and four next-to-nearest-neighbor 
spins ($t_1^\prime,t_2^\prime,t_3^\prime,t_4^\prime$),  
which are connected to $t_0$ through the 
plaquette interaction.   
We associate a different value of $c_{0,j}$ to each type  
of spin ($c_{0,{\rm nn}}$ and $c_{0,{\rm nnn}}$, respectively).  
The quantity $\alpha$ is now equal to
$\alpha = 2 c_{0,{\rm nn}} + 4 c_{0,{\rm nnn}}$.

As explained in the last subsection, we have to include in the statistical 
weight the delta functions corresponding to the surviving original links and 
involving the spin $t_0$. In this example there are two such factors:  
$(1 - \delta_{t_0,t_1})(1 - \delta_{t_0,t_2})$.  
The numerical results for $q=5,6,7,8$ are 
displayed in Table~\ref{Table_results_kagome}.  
The configurations which minimize $c_{0,j}$ for each type of neighbor $t_j$
are shown in Table~\ref{Table_configurations_kagome}. 
Using the configurations depicted in Table~\ref{Table_configurations_kagome}, 
it is easy to guess the general formulae for $c_{0,{\rm nn}}$ and 
$c_{0,{\rm nnn}}$:  
\begin{eqnarray}
c_{0,{\rm nn}} &=& { q-3 \over
              q^2 - 7q +13 } \; 
 \quad \hbox{for } q \geq 5 \\[2mm] 
c_{0,{\rm nnn}} &=& { q-3 \over
              q^3 - 10 q^2 + 35 q - 43 } 
 \quad \hbox{for } q \geq 5  
\end{eqnarray}

We see that Dobrushin's condition $\alpha<1$ holds for $q\geq7$. 
Again we obtained an improvement of two units over Koteck\'y's result 
($q\geq9$). Our 
bound $q\geq7$ should be compared to the exact result $q\geq4$ (or more 
precisely, $q>3$).

%
%
\section{Further Improvements: Cluster Decimation} \label{sec5}

In this section we present slightly better results for the hexagonal and 
Kagom\'e lattices. The idea is simple: if using single-site decimation 
(Section~\ref{sec4}) we obtained improved bounds, then it is natural to 
expect even better results if we decimate clusters of nearby spins. 
This is what happens in the proof presented in Ref.~\cite{Kennedy_95}, and 
it happens 
also in our case. Obviously, as we decimate larger clusters, the 
effective interaction among the remaining spins becomes more and more 
complicated (the effective interaction contains between 128 and 2410 terms 
for the three cases considered below); 
this fact limits the practical utility of this method. 
Nevertheless, we have been able to improve slightly our previous  
results in two cases: 
the hexagonal and Kagom\'e lattices.    
Our method will be explained in detail in the following subsections.

\subsection{Hexagonal lattice}

The first step is to choose suitable clusters of spins to be summed over. 
In this example we selected a subset of the hexagonal faces of the original 
lattice (see Figure~\ref{figure_hexagonal_lattice_bis}).  
The remaining spins (solid circles) define the decimated lattice, which is 
again a hexagonal lattice. Each hexagonal face of this  
decimated lattice contains one hexagonal cluster of spins that were 
summed over (empty circles).    
It is important to notice that these clusters (of empty circles)  
do not have any nearest-neighbor interactions with other such  
clusters. So, we can perform the sum over the six spins belonging to the  
cluster, and obtain an effective 
interaction among the six spins of the decimated lattice  
surrounding the cluster.   
However, this effective interaction is not as simple as the 
single-site-decimation interaction \reff{decimation_factor}.  
To be able to handle it, we wrote a program in {\sc Mathematica} to do all 
the required sums. The final expression can be written as a certain 
linear combination of products of Kronecker delta functions. This turns to be 
very long and complicated, 
so we omit its form here.\footnote{ 
  Actually, we did not use this expression in our {\sc Fortran} programs, as 
  it is very memory- and time-consuming. 
  Rather, we first classified all the possible  
  configurations into classes with the 
  same statistical weight. For the cases considered here, the number of classes 
  is moderate (up to 36). The important point is that we can  
  easily tell to which 
  class a given configuration belongs, by measuring a few quantities 
  (such as, for instance, the number of distinct colors of the configuration).  
  We then devise a simple formula that reproduces the correct statistical 
  weight. 
  The practical procedure depends 
  on the lattice and type of decimation considered, but it is always 
  faster and less memory-consuming than direct use of the formula  
  computed with {\sc Mathematica}.    
}    
We remark that  
this interaction has the property that even when $q=3=r$, every state of $t_0$ 
gets nonzero weight, irrespective of the configuration of the   
neighboring spins; in particular, there is no  ambiguity at $T=0$, in 
contrast to what happens for the  
single-site-decimation interaction \reff{decimation_factor}.

The effective interaction lives on the hexagonal faces of  
the decimated lattice. 
Each spin $t_0$ interacts with three hexagons: 
$(t_0,t_1,t_1^\prime,t_1^{\prime\prime},t_2^\prime,t_2)$, 
$(t_0,t_2,t_3^\prime,t_2^{\prime\prime},t_4^\prime,t_3)$, and  
$(t_0,t_3,t_5^\prime,t_3^{\prime\prime},t_6^\prime,t_1)$. There are three 
types of neighbors: three nearest-neighbor spins $t_1,t_2,t_3$ (which 
belong to two different hexagons), 
six second-nearest-neighbor spins $t_1^\prime,\ldots,t_6^\prime$ (which belong 
to only one hexagon and which are not diametrically 
opposite to $t_0$), and three third-neighbor spins  
$t_1^{\prime\prime},t_2^{\prime\prime},t_3^{\prime\prime}$ 
(which belong to only one hexagon and which are diametrically  
opposite to $t_0$). We have to compute a different $c_{0,j}$ for each type 
of neighbor: we denote these $c_{0,1{\rm n}}$, $c_{0,2{\rm n}}$ 
and $c_{0,3{\rm n}}$,  
respectively. The quantity \reff{def0_alpha} is now equal to 
$\alpha= 3c_{0,1{\rm n}} + 6 c_{0,2{\rm n}} + 3c_{0,3{\rm n}}$.

The numerical results for $q=3,4$ are displayed in  
Table~\ref{Table_results_hexagonal_bis}. We see that Dobrushin's condition 
$\alpha<1$ is satisfied for $q=4$, so we have improved by one unit the 
bound of Section~\ref{sec4}. That is, we have proven that  
for the hexagonal lattice at zero temperature there is exponential decay of
correlations for 
$q\geq4$. The expected result is $q> 2.618\ldots\;$.

\subsection{Kagom\'e lattice}

In this case, our chosen clusters will be a subset of the triangular faces 
of the Kagom\'e lattice (empty circles in 
Figure~\ref{figure_kagome_lattice_bis}); they are not connected by any  
nearest-neighbor interaction.  The remaining spins (solid circles)  
define the decimated lattice, which turns out to be hexagonal. The triangular 
clusters are surrounded by the ``deformed'' hexagonal faces of the 
decimated lattice. In addition, there are hexagonal faces of the original 
lattice which belong also to the decimated lattice. The effective interaction 
coming from the decimation procedure 
lives on the ``deformed''  hexagonal faces only.   
It has again a very complicated form, and we had to
use {\sc Mathematica} to compute it. When $q=4=r$, we see that 
the effective interaction assigns a zero weight to a few configurations. 
However, there are no configurations of the   
neighboring spins for which {\em all} the possible values of $t_0$ are 
forbidden, so there is no ambiguity at $T=0$ even in this case.

{}From Figure~\ref{figure_kagome_lattice_bis} we see that each spin $t_0$ of 
the decimated lattice interacts with only two ``deformed'' hexagons. 
There are 
four types of neighboring spins: two nearest-neighbor spins $t_1,t_2$ 
(which belong to only one hexagon), one 
second-nearest neighbor $t_1^\prime$ (which belongs to both hexagons), 
four third-nearest neighbors 
$t_1^{\prime\prime},t_2^{\prime\prime},t_3^{\prime\prime},t_4^{\prime\prime}$  
(which belong to one hexagon and are not connected to $t_1^\prime$), 
and two fourth-nearest neighbors 
$t_1^{\prime\prime\prime},t_2^{\prime\prime\prime}$ (which belong to only 
one hexagon and are connected to $t_1^\prime$).   
Notice that, in addition to the 
effective interaction coming from the decimation procedure, we 
have to include the factors $(1-\delta_{t_0,t_1})(1-\delta_{t_0,t_2})$ arising 
from the original 2-body interaction, 
because the links $\<t_0,t_1\>$ and $\<t_0,t_2\>$ belong also to the 
original lattice. To each neighbor type we associate a different $c_{0,j}$:   
we denote these 
$c_{0,1{\rm n}}$, $c_{0,2{\rm n}}$, $c_{0,3{\rm n}}$ and $c_{0,4{\rm n}}$,  
respectively. The quantity \reff{def0_alpha} takes the form 
$\alpha= 2c_{0,1{\rm n}}+c_{0,2{\rm n}}+4c_{0,3{\rm n}}+2c_{0,4{\rm n}}$.

In Table~\ref{Table_results_kagome_bis} we show our numerical results 
for $q=4,5,6$. We notice that the constants for the third and fourth 
nearest neighbors coincide in all cases. 
(However, we were unable to find an analytic proof of this result. In 
particular, there does not appear to be any symmetry that would yield 
this equality.)  
Dobrushin's condition $\alpha<1$ 
holds for $q=6$, improving the result of Section~\ref{sec4} by one unit. 
So, there is no phase transition at zero temperature for the Kagom\'e 
lattice when $q\geq6$, which should be compared with the expected result 
$q>3$.

We have also tried to improve these results by considering decimation of 
hexagonal clusters (as we did in the previous subsection). After the 
decimation procedure we obtained a new Kagom\'e lattice (rotated 90 degrees). 
The values of $\alpha$ for $q=4,5$ were 3.83 and 1.07, respectively, which are 
smaller than the corresponding values reported in  
Table~\ref{Table_results_kagome_bis}. 
However, in both cases $\alpha>1$. Therefore,  
we are unable to prove exponential 
decay of correlations for $q<6$.

%
%
\section*{Acknowledgments}

We wish to thank Chris Henley for sending us drafts of his work  
\cite{Henley_94} 
prior to publication.  

The authors' research was supported in part by
a M.E.C.\  (Spain)/Fulbright fellowship (J.S.),
and by U.S.\ National Science Foundation grants DMS-9200719 
and PHY-9520978 (J.S.\ and A.D.S.).

\newpage 
\renewcommand{\baselinestretch}{1}
\large\normalsize
%
%
%
%

\clearpage

\newpage 

%
%
\def\kk{\phantom{1}}
\def\mm{\phantom{.618\ldots}} 
\begin{table}[htb]
\centering
\begin{tabular}{l|l|c|c|c|c|}
\cline{2-6} \cline{2-6} \\[-0.5cm]
            & \multicolumn{5}{c|}{Range of $q$}  \\  
\cline{2-6} \\[-0.5cm] 
            & \multicolumn{1}{|c|}{General}   
            & \multicolumn{1}{|c|}{Hexagonal}  
            & Square & Kagom\'e & Triangular \\[0.1cm]
\hline \hline 
\multicolumn{1}{|l|}{Koteck\'y}& 
  $> 2r$           & $\geq 7$\mm & $\geq9$ & $\geq9$ & $\geq13$    \\  
\multicolumn{1}{|l|}{Single-site Decimation}& 
                   & $\geq5$\mm  & $\geq7$ & $\geq7$ & $\geq11$    \\  
\multicolumn{1}{|l|}{Cluster Decimation}&
                   & $\geq4$\mm  &         & $\geq6$ &       \\  
\hline    
\multicolumn{1}{|l|}{Exact}    &  
                   & $>2.618\ldots$  
                                 & $>3$    & $>3$    & $>4\kk$  \\  
\hline\hline
\end{tabular}
\caption{ 
Range of $q$ for which we have proven exponential decay of correlations at  
zero temperature for various lattices.   
The first row shows the result given 
by Koteck\'y \protect\cite[pp.~148--149, 457]{Georgii_88}   
and slightly generalized here in 
Section~\protect\ref{sec3}.  
The second row gives our improved result using single-site decimation  
(Section~\protect\ref{sec4}), and 
the next row gives our further improvement using more sophisticated 
decimation schemes (Section~\protect\ref{sec5}).  
The last row (``Exact") shows what is known  
or believed to be the right answer.
}
\label{Table_results}
\end{table}

%
%
\begin{table}[htb] 
\centering 
\begin{tabular}{|c|c|c|l|}
\hline \hline \\[-0.5cm]
$q$ & $c_{0,{\rm nn}}$ & $c_{0,{\rm nnn}}$ &  
      \multicolumn{1}{|c|}{$\alpha$} \\[0.1cm]
\hline \hline
5 & 0.3750  & 0.2353 & 2.4412 \\ 
6 & 0.1899  & 0.1093 & 1.1967 \\ 
7 & 0.1137  & 0.0636 & 0.7093 $< 1$ \\ 
8 & 0.0756  & 0.0415 & 0.4683 $< 1$ \\ 
\hline\hline
\end{tabular}
\caption{Numerical results for the square lattice. For each value of $q$ we 
show the quantities $c_{0,{\rm nn}}$ and $c_{0,{\rm nnn}}$.    
Finally, we give the value of the parameter  
$\alpha = 4c_{0,{\rm nn}}+4c_{0,{\rm nnn}}$. When $\alpha<1$   
there is a unique Gibbs measure at $T=0$.  
}
\label{Table_results_square} 
\end{table}  
 
%
%
\newcommand{\square}[8]{ 
   \begin{picture}(100,70)
      \put(20,10){#8} 
      \put(20,35){#5} 
      \put(20,60){#3} 
      \put(45,10){#7} 
      \put(47,36){$\circ$}
      \put(45,60){#2} 
      \put(70,10){#6}
      \put(70,35){#4} 
      \put(70,60){#1} 
      \multiput(32,14)(0,25){3}{\line(1,0){11}} 
      \multiput(57,14)(0,25){3}{\line(1,0){11}} 
      \multiput(25,21)(25,0){3}{\line(0,1){11}} 
      \multiput(25,46)(25,0){3}{\line(0,1){11}} 
   \end{picture} 
} 

\begin{table}[htb]
\centering
\begin{tabular}{|cc|c|}
\hline \hline \\[-0.5cm]
\multicolumn{2}{|c|}{$c_{0,{\rm nn}}$} & $c_{0,{\rm nnn}}$ \\[0.1cm]
\hline \hline
    &    &    \\
\square{A}{B}{C}{C,D}{D}{A}{E}{C} & 
\square{A}{B}{C}{C,D}{D}{F}{E}{C} & 
\square{A,B}{C}{A}{D}{B}{B}{A}{E} \\  
$q=5$ & $q \geq 6$ & $q \geq 5$ \\
\hline\hline
\end{tabular}
\caption{Configurations which maximize $c_{0,{\rm nn}}$ and 
$c_{0,{\rm nnn}}$ for the square lattice when $q\geq 5$. Each distinct   
letter represents a distinct spin value.  
The spin $t_0$ is denoted by an empty circle ($\circ$).     
The spin $t_j$ is the one that has two different 
spin values associated to it.  
}
\label{Table_configurations_square} 
\end{table} 

%
%
\begin{table}[htb]
\centering
\begin{tabular}{|c|c|l|}
\hline \hline \\[-0.5cm]
$q$ & $c_{0,{\rm nn}}$ &  \multicolumn{1}{|c|}{$\alpha$} \\[0.1cm]
\hline \hline
4 & 0.2667  & 1.6000 \\
5 & 0.1233  & 0.7397 $< 1$ \\
6 & 0.0699  & 0.4192 $< 1$ \\
\hline\hline
\end{tabular}
\caption{Numerical results for the hexagonal lattice. For each value of $q$ we
compute the quantities $c_{0,{\rm nn}}$ and $\alpha=6c_{0,{\rm nn}}$.  
}
\label{Table_results_hexagonal}
\end{table}

%
%
\newcommand{\hexagon}[6]{ 
   \setlength{\unitlength}{0.33pt} 
   \begin{picture}(300,200)
      \put(170,183.2){#1} 
      \put(70,183.2){#2} 
      \put(20,96.6){#3} 
      \put(70,10){#4} 
      \put(215,96.6){#5} 
      \put(170,10){#6} 
      \put(117,96.6){$\circ$}

      \put(107,192.2){\line(1,0){50}} 
      \put(57,105.2){\line(1,0){50}} 
      \put(147,105.2){\line(1,0){50}} 
     
      \put(87,40){\line(3,5){30}} 
      \put(187,40){\line(3,5){30}} 
      \put(137,121.6){\line(3,5){30}} 

      \put(63,40){\line(-3,5){30}} 
      \put(163,40){\line(-3,5){30}} 
      \put(113,121.6){\line(-3,5){30}} 

   \end{picture} 
} 

\begin{table}[htb]
\centering
\begin{tabular}{|c|}
\hline \hline \\[-0.5cm]
$c_{0,{\rm nn}}$ \\[0.1cm]
\hline \hline
\\ 
\hexagon{A}{B}{A}{B}{A,B}{C} \\  
$q \geq 4$  \\
\hline\hline
\end{tabular}
\caption{Configurations which maximize $c_{0,{\rm nn}}$ for
the hexagonal lattice when $q\geq4$. The notation is as 
in Table~\protect\ref{Table_configurations_square}. 
Notice that, because the two outer spins within each triangle  
are equivalent, we can freely permute their values.  
}
\label{Table_configurations_hexagon} 
\end{table} 

%
%
\begin{table}[htb]
\centering
\begin{tabular}{|r|c|c|l|}
\hline \hline \\[-0.5cm]
$q$ & $c_{0,{\rm nn}}$ & $c_{0,{\rm nnn}}$ &  
      \multicolumn{1}{|c|}{$\alpha$} \\[0.1cm]
\hline \hline
7 & 0.5714  &  0.2667 & 4.1143 \\
8 & 0.3803  &  0.1233 & 2.2504 \\
9 & 0.2832  &  0.0699 & 1.4784 \\
10& 0.2244  &  0.0446 & 1.0743 \\ 
11& 0.1852  &  0.0307 & 0.8324 $<1$ \\ 
12& 0.1574  &  0.0224 & 0.6741 $<1$ \\ 
\hline\hline
\end{tabular}
\caption{Numerical results for the triangular lattice. Notation is as in 
Table~\protect\ref{Table_results_square};  
here $\alpha=3c_{0,{\rm nn}}+9c_{0,{\rm nnn}}$.  
}
\label{Table_results_triangular}
\end{table}

%
%
\newcommand{\triangular}[7]{ 
   \setlength{\unitlength}{0.33pt} 
   \begin{picture}(500,400)
      \put(220,356.4){C} 
      \put(320,356.4){B} 
      \put(70,269.8){E} 
      \put(170,269.8){D} 
      \put(370,269.8){#1} 
      \put(20,183.2){F} 
      \put(217,183.2){$\circ$}
      \put(320,183.2){#2} 
      \put(70,96.6){#3}
      \put(170,96.6){#4}
      \put(370,96.6){#5}
      \put(220,10){#6}
      \put(320,10){#7}

      \put(257,365.4){\line(1,0){50}} 
      \put(107,278.8){\line(1,0){50}} 
      \put(257,192.2){\line(1,0){50}} 
      \put(107,105.6){\line(1,0){50}} 
      \put(257,19){\line(1,0){50}} 
     
      \put(37,213.2){\line(3,5){30}} 
      \put(187,299.8){\line(3,5){30}} 
      \put(187,126.6){\line(3,5){30}} 
      \put(337,40){\line(3,5){30}} 
      \put(337,213.2){\line(3,5){30}} 

      \put(63,126.6){\line(-3,5){30}} 
      \put(213,40){\line(-3,5){30}} 
      \put(213,213.2){\line(-3,5){30}} 
      \put(363,126.6){\line(-3,5){30}} 
      \put(363,299.8){\line(-3,5){30}} 

   \end{picture} 
} 

\begin{table}[htb]
\centering
\begin{tabular}{|c|c|}
\hline \hline \\[-0.5cm]
$c_{0,{\rm nn}}$  & $c_{0,{\rm nnn}}$ \\[0.1cm]
\hline \hline 
\triangular{A}{C,E}{G}{B}{D}{G}{C} &  
\triangular{E,F}{A}{A}{C}{D}{F}{E} \\  
\triangular{A}{C,E}{C}{A}{D}{G}{C} &  \\ 
$q \geq 7$   & $q \geq 7$  \\
\hline\hline
\end{tabular}
\caption{Configurations maximizing $c_{0,{\rm nn}}$ and $c_{0,{\rm nnn}}$ 
for the triangular lattice when $q\geq 7$. Notation is as in  
Table~\protect\ref{Table_configurations_square}. Notice that, as in 
Table~\protect\ref{Table_configurations_hexagon}, we obtain equivalent 
configurations by permuting the outer spins within each hexagon (i.e.\  
those spins that are not shared between two hexagons). 
}
\label{Table_configurations_triangular} 
\end{table}

%
%
\begin{table}[htb]
\centering
\begin{tabular}{|r|c|c|l|}
\hline \hline \\[-0.5cm]
$q$ & $c_{0,{\rm nn}}$ & $c_{0,{\rm nnn}}$ &  
      \multicolumn{1}{|c|}{$\alpha$} \\[0.1cm]
\hline \hline
5 & 0.6667  &  0.2857 & 2.4762 \\
6 & 0.4286  &  0.1304 & 1.3789 \\
7 & 0.3077  &  0.0727 & 0.9063 $<1$ \\
8 & 0.2381  &  0.0459 & 0.6597 $<1$  \\
\hline\hline
\end{tabular}
\caption{Numerical results for the Kagom\'e lattice. Notation is as in 
Table~\protect\ref{Table_results_square}; here 
$\alpha=2c_{0,{\rm nn}}+4c_{0,{\rm nnn}}$.  
}
\label{Table_results_kagome}
\end{table}

%
%
\newcommand{\kagome}[6]{
   \begin{picture}(100,70)
      \put(20,10){#5}
      \put(20,35){#3}
      \put(45,10){#6}
      \put(47,36){$\circ$}
      \put(45,60){#1}
      \put(70,35){#4}
      \put(70,60){#2}
      \multiput(32,14)(0,25){2}{\line(1,0){11}}
      \multiput(57,39)(0,25){2}{\line(1,0){11}}
      \multiput(25,21)(25,0){2}{\line(0,1){11}}
      \multiput(50,46)(25,0){2}{\line(0,1){11}}
      \thicklines 
      \multiput(57,39)(-25,0){2}{\line(1,0){11}} 
   \end{picture}
}
\begin{table}[htb]
\centering
\begin{tabular}{|c|c|}
\hline \hline \\[-0.5cm]
$c_{0,{\rm nn}}$ & $c_{0,{\rm nnn}}$ \\[0.1cm]
\hline \hline
    &       \\
\kagome{B}{A}{B}{A,C}{C}{A} & \kagome{B}{C,D}{B}{A}{C}{D} \\  
 $q \geq 5$ & $q \geq 5$ \\
\hline\hline
\end{tabular}
\caption{Configurations which maximize $c_{0,{\rm nn}}$ and $c_{0,{\rm nnn}}$ 
for the Kagom\'e lattice when $q\geq5$. Notation is as in 
Table~\protect\ref{Table_configurations_square}.  
In the pictures, a thick line represents a link belonging to the  
original lattice (and carrying an additional Kronecker delta term). 
On each square, the two spins not connected to such a link 
can be freely interchanged.  
}
\label{Table_configurations_kagome}
\end{table} 

%
%
\begin{table}[htb]
\centering
\begin{tabular}{|r|c|c|c|l|}
\hline \hline \\[-0.5cm]
$q$ & $c_{0,1{\rm n}}$ & $c_{0,2{\rm n}}$ & $c_{0,3{\rm n}}$ &
      \multicolumn{1}{|c|}{$\alpha$} \\[0.1cm]
\hline \hline
3 & 0.5685  &  0.3904 & 0.2136 & 4.5985 \\
4 & 0.1036  &  0.0292 & 0.0164 & 0.5356 $<1$ \\
\hline\hline
\end{tabular}
\caption{Numerical results for the hexagonal lattice when we use 6-spin 
decimation. For each type of neighbor we show the corresponding value  
$c_{0,j}$.  
We also show $\alpha=3c_{0,1{\rm n}} + 6 c_{0,2{\rm n}} + 3c_{0,3{\rm n}}$. 
When $\alpha<1$ there is a unique Gibbs measure at $T=0$. 
}
\label{Table_results_hexagonal_bis}
\end{table}

%
%
\begin{table}[htb]
\centering
\begin{tabular}{|r|c|c|c|c|l|}
\hline \hline \\[-0.5cm]
$q$ & $c_{0,1{\rm n}}$ & $c_{0,2{\rm n}}$ & $c_{0,3{\rm n}}$  
    & $c_{0,4{\rm n}}$ & \multicolumn{1}{|c|}{$\alpha$} \\[0.1cm]  
\hline \hline
4 & 1.0000  &  1.0000 & 0.6667 & 0.6667 & 7.0000 \\
5 & 0.4949  &  0.1590 & 0.1003 & 0.1003 & 1.7504 \\
6 & 0.2975  &  0.0581 & 0.0330 & 0.0330 & 0.8529 $<1$ \\
\hline\hline
\end{tabular}
\caption{Numerical results for the Kagom\'e lattice when we use 3-spin
decimation. Notation is as in Table~\protect\ref{Table_results_hexagonal_bis}; 
here $\alpha=2c_{0,1{\rm n}}+c_{0,2{\rm n}}+4c_{0,3{\rm n}}+2c_{0,4{\rm n}}$.  
}
\label{Table_results_kagome_bis}
\end{table}


\newpage 


%
%

\newpage 

%
%

\newpage

%
%

\newpage 

%
%
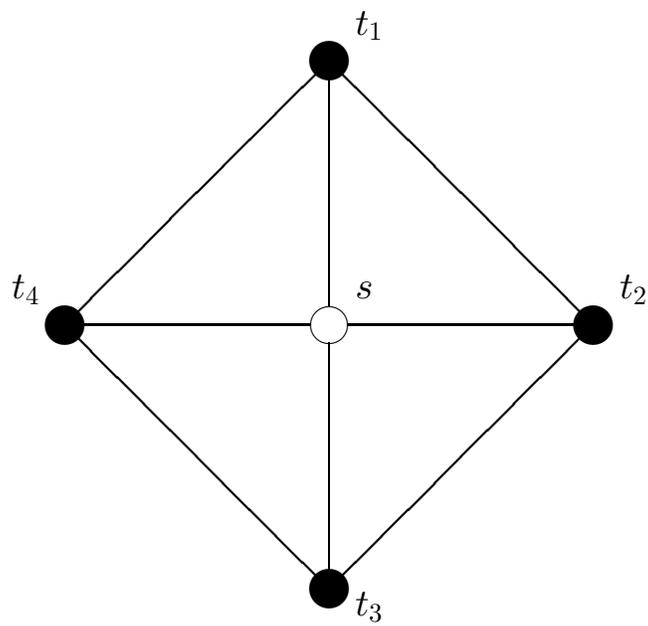
\begin{figure}[t] 
\centering 
\begin{picture}(400,400)(0,0) 
     \put(200,200){\circle{14}}  
     \put(210,210){\large $s$} 
     \put(200,300){\circle*{20}}  
     \put(210,310){\large $t_1$} 
     \put(200,100){\circle*{20}}  
     \put(210,90){\large $t_3$} 
     \put(300,200){\circle*{20}}  
     \put(310,210){\large $t_2$} 
     \put(100,200){\circle*{20}}  
     \put(80,210){\large $t_4$} 
     \thinlines 
     \put(200,207){\line(0,1){93}} 
     \put(200,193){\line(0,-1){93}} 
     \put(207,200){\line(1,0){93}} 
     \put(193,200){\line(-1,0){93}} 
     \thicklines
     \put(200,300){\line(1,-1){100}} 
     \put(200,300){\line(-1,-1){100}} 
     \put(200,100){\line(1,1){100}} 
     \put(200,100){\line(-1,1){100}} 
\end{picture}
\caption{Decimation for the square-lattice case. Once we sum over the spin 
$s$, we obtain a new effective interaction among the spins $t_j$  
($j=1,2,3,4$).} 
\label{figure_decimation_square}
\end{figure}  

\newpage 

%
%
%

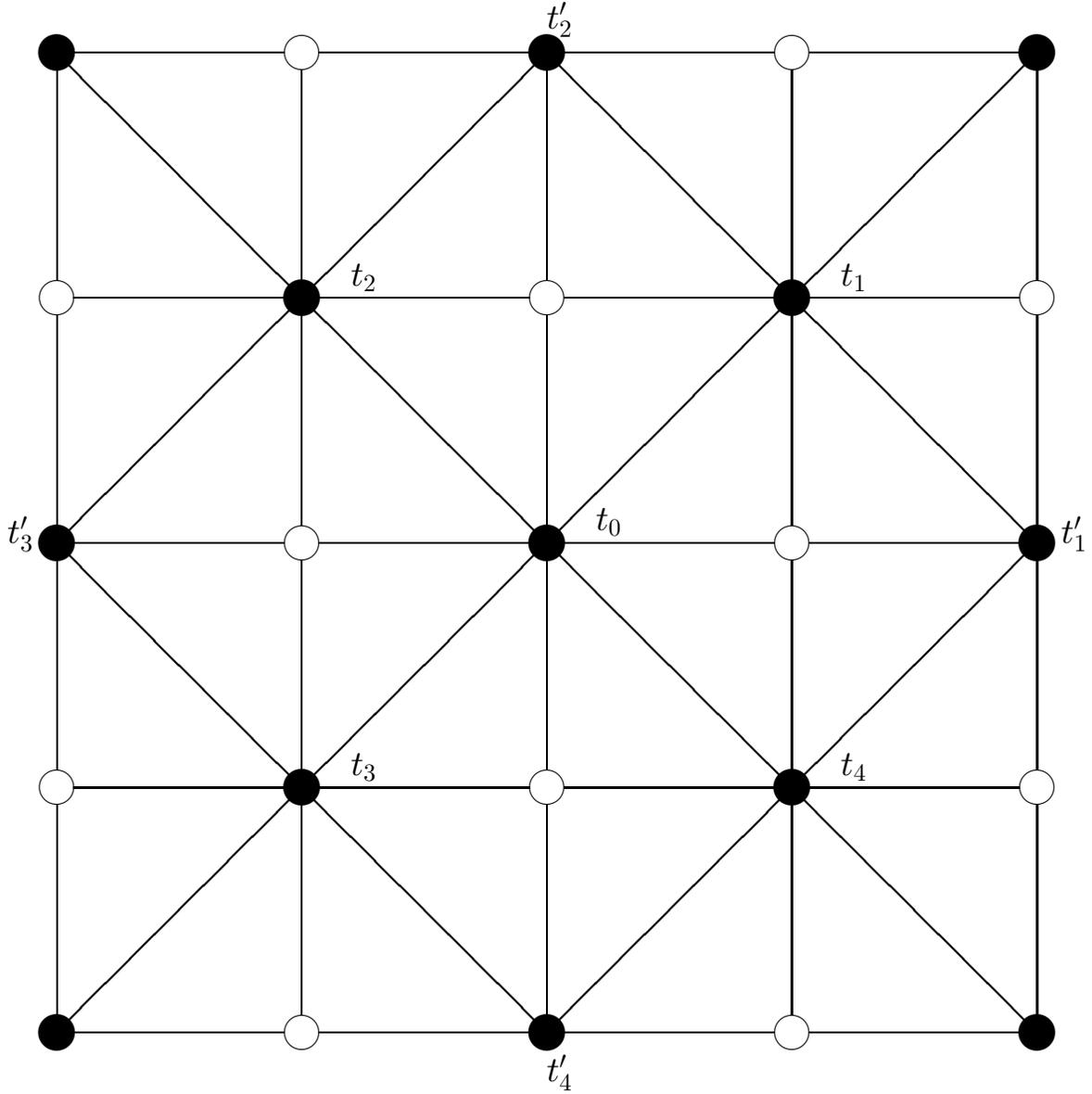
\begin{figure}
\centering 
\begin{picture}(400,400)(0,0) 

\thinlines
\newsavebox{\hor}
\savebox{\hor}(100,100)[bl]{\begin{picture}(100,100)
\put(7,0){\line(1,0){86}} 
\end{picture}}   

\newsavebox{\ver}
\savebox{\ver}(100,100)[bl]{\begin{picture}(100,100)
\put(0,7){\line(0,1){86}}
\end{picture}} 

\multiput(0,0)(100,0){4}{\usebox{\hor}} 
\multiput(0,100)(100,0){4}{\usebox{\hor}} 
\multiput(0,200)(100,0){4}{\usebox{\hor}} 
\multiput(0,300)(100,0){4}{\usebox{\hor}} 
\multiput(0,400)(100,0){4}{\usebox{\hor}} 

\multiput(0,0)(0,100){4}{\usebox{\ver}} 
\multiput(100,0)(0,100){4}{\usebox{\ver}} 
\multiput(200,0)(0,100){4}{\usebox{\ver}} 
\multiput(300,0)(0,100){4}{\usebox{\ver}} 
\multiput(400,0)(0,100){4}{\usebox{\ver}} 
 
\multiput(0,0)(200,0){3}{\circle*{20}} 
\multiput(0,200)(200,0){3}{\circle*{20}} 
\multiput(0,400)(200,0){3}{\circle*{20}} 
\multiput(100,100)(200,0){2}{\circle*{20}} 
\multiput(100,300)(200,0){2}{\circle*{20}} 

\multiput(0,100)(200,0){3}{\circle{14}}
\multiput(0,300)(200,0){3}{\circle{14}}
\multiput(100,0)(200,0){2}{\circle{14}}
\multiput(100,200)(200,0){2}{\circle{14}}
\multiput(100,400)(200,0){2}{\circle{14}} 

\thicklines
\put(0,200){\line(1,1){200}}
\put(0,0){\line(1,1){400}}
\put(200,0){\line(1,1){200}}
\put(0,200){\line(1,-1){200}}
\put(0,400){\line(1,-1){400}}
\put(200,400){\line(1,-1){200}}

\put(220,205){\large $t_0$}
\put(410,200){\large $t_1^\prime$}
\put(200,410){\large $t_2^\prime$}
\put(200,-20){\large $t_4^\prime$}
\put(-20,200){\large $t_3^\prime$} 
\put(320,305){\large $t_1$}
\put(120,305){\large $t_2$}
\put(120,105){\large $t_3$}
\put(320,105){\large $t_4$}

\end{picture}
\vspace*{1cm}
\caption{Decimation for the square lattice. The empty circles represent the 
spins summed over; the solid circles represent the resulting decimated lattice. 
Each spin $t_0$ of the decimated lattice interacts with 
eight spins: four nearest neighbors $t_1,t_2,t_3,t_4$ and four  
next-to-nearest neighbors $t_1^\prime,t_2^\prime,t_3^\prime,t_4^\prime$. 
} 
\label{figure_square_lattice}
\end{figure}  

\newpage 

%
%
%

\begin{figure}
\setlength{\unitlength}{0.75pt} 
\begin{picture}(500,500)(0,0) 

\thinlines
\newsavebox{\horhex}
\savebox{\horhex}(100,100)[bl]{\begin{picture}(100,100)
\put(7,0){\line(1,0){86}} 
\end{picture}}   

\multiput(50,0)(300,0){2}{\usebox{\horhex}} 
\multiput(50,173.2)(300,0){2}{\usebox{\horhex}} 
\multiput(50,346.4)(300,0){2}{\usebox{\horhex}} 

\multiput(200,86.6)(0,173.2){3}{\usebox{\horhex}}

\multiput(50,0)(300,0){2}{\circle*{15}} 
\multiput(50,173.2)(300,0){2}{\circle*{15}} 
\multiput(50,346.4)(300,0){2}{\circle*{15}} 
\multiput(150,0)(300,0){2}{\circle{14}}
\multiput(150,173.2)(300,0){2}{\circle{14}}
\multiput(150,346.4)(300,0){2}{\circle{14}} 

\multiput(200,86.6)(300,0){2}{\circle*{15}}
\multiput(200,259.8)(300,0){2}{\circle*{15}}
\multiput(200,433.0)(300,0){2}{\circle*{15}}
\multiput(0,86.6)(300,0){2}{\circle{14}}
\multiput(0,259.8)(300,0){2}{\circle{14}}
\multiput(0,433.0)(300,0){2}{\circle{14}} 

\newsavebox{\diagup}
\savebox{\diagup}(100,100)[bl]{\begin{picture}(100,100)
\put(3.5,6.062){\line(3,5){44}}
\end{picture}} 

\multiput(150,0)(-150,86.6){2}{\usebox{\diagup}}
\multiput(450,173.2)(-150,86.6){3}{\usebox{\diagup}}
\multiput(450,0)(-150,86.6){4}{\usebox{\diagup}}
\put(450,346.4){\usebox{\diagup}}

\newsavebox{\diagdown}
\savebox{\diagdown}(100,100)[bl]{\begin{picture}(100,100)
\put(3.5,-6.062){\line(3,-5){44}}
\end{picture}} 

\multiput(0,86.6)(150,86.6){4}{\usebox{\diagdown}} 
\multiput(300,86.6)(150,86.6){2}{\usebox{\diagdown}} 
\multiput(0,259.8)(150,86.6){3}{\usebox{\diagdown}} 
\put(0,433.0){\usebox{\diagdown}} 

\thicklines
\put(50,0){\line(0,1){433}} 
\put(200,0){\line(0,1){433}} 
\put(350,0){\line(0,1){433}} 
\put(500,0){\line(0,1){433}}

\put(56,171){\line(5,-3){144}} 
\put(56,344.2){\line(5,-3){144}} 
\put(206,257.6){\line(5,-3){144}} 
\put(206,430.8){\line(5,-3){144}} 
\put(206,84.4){\line(5,-3){144}} 
\put(356,171){\line(5,-3){144}} 
\put(356,344.4){\line(5,-3){144}} 

\put(56,349){\line(5,3){144}} 
\put(56,175.8){\line(5,3){144}} 
\put(56,2.6){\line(5,3){144}} 
\put(206,262.4){\line(5,3){144}} 
\put(206,89.2){\line(5,3){144}} 
\put(356,2.6){\line(5,3){144}} 
\put(356,175.8){\line(5,3){144}} 
\put(356,349){\line(5,3){144}} 

\put(50,0){\line(-5,3){50}} 
\put(50,173.2){\line(-5,-3){50}} 
\put(50,173.2){\line(-5,3){50}} 
\put(50,346.4){\line(-5,3){50}} 
\put(50,346.4){\line(-5,-3){50}} 

\put(360,143){\large $t_1$}  
\put(360,316){\large $t_2$}  
\put(210,403){\large $t_3$}  
\put(60,316){\large $t_4$}  
\put(60,143){\large $t_5$}  
\put(210,57){\large $t_6$}  
\put(210,230){\large $t_0$}  

\end{picture}
\vspace*{1cm}
\caption{Decimation for the hexagonal lattice. The original hexagonal lattice 
is drawn with thin lines; 
the empty circles represent the spins summed over;  
the triangular lattice resulting from decimation
is drawn with solid circles and thick lines.  
Each spin $t_0$ of the decimated lattice interacts with  
six nearest-neighbor spins $t_1,\ldots,t_6$.   
} 
\label{figure_hexagonal_lattice}
\end{figure}
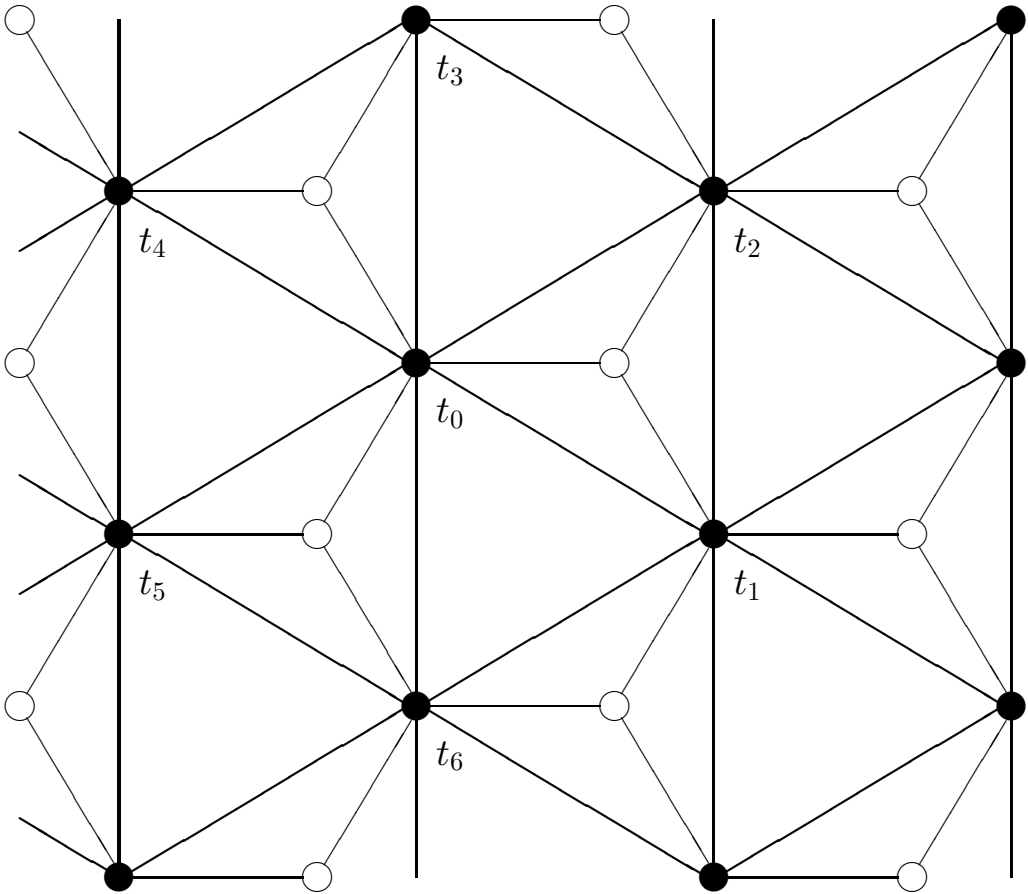  

\newpage 

%
%
%

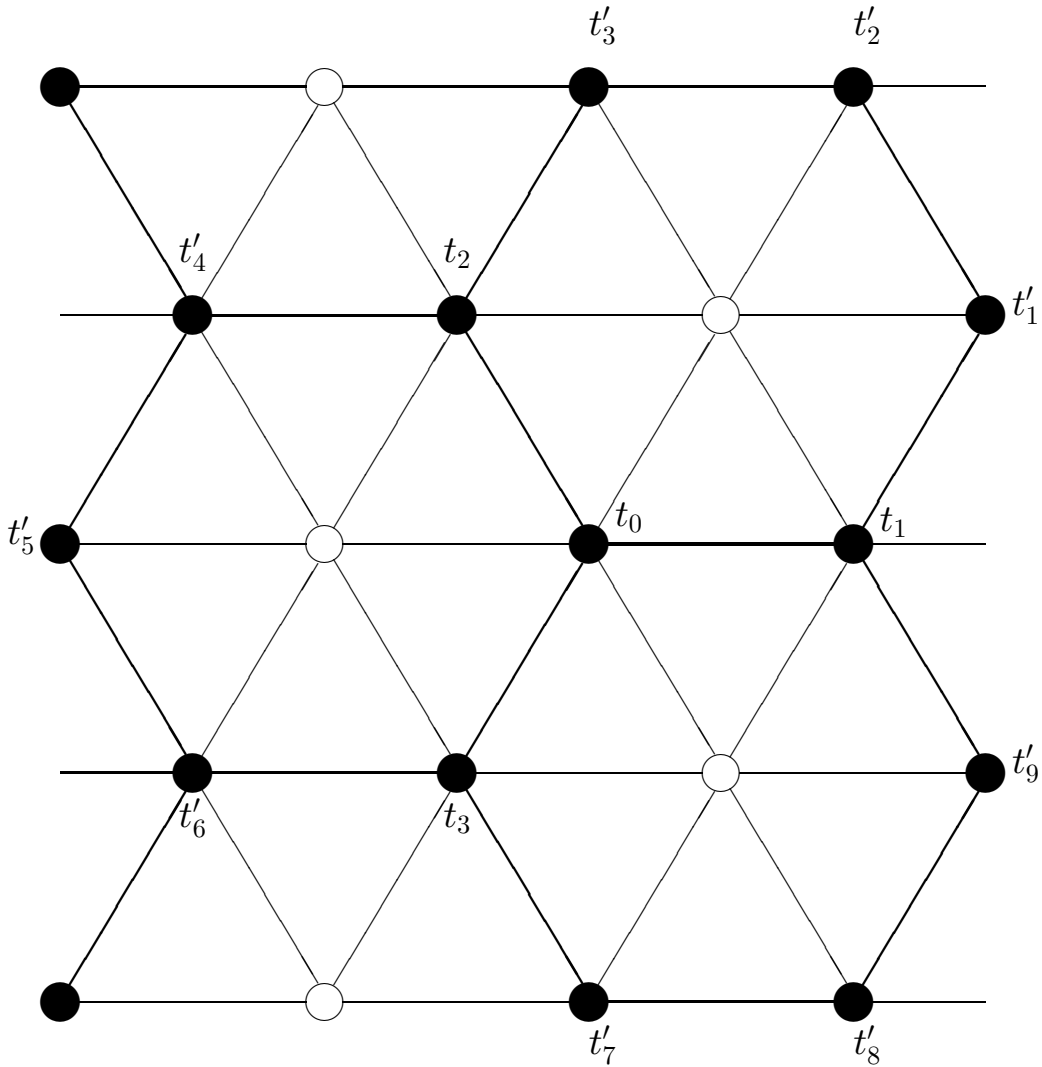
\begin{figure}
\centering 
\begin{picture}(350,350)(0,0) 

\thinlines

\newsavebox{\hortri}
\savebox{\hortri}(200,100)[bl]{\begin{picture}(200,100)
\put(0,0){\line(1,0){93}}
\put(200,0){\line(-1,0){93}} 
\end{picture}}

\multiput(0,0)(0,173.2){3}{\usebox{\hortri}} 
\multiput(150,86.6)(0,173.2){2}{\usebox{\hortri}} 
\multiput(200,0)(0,173.2){3}{\line(1,0){150}} 
\multiput(0,86.6)(0,173.2){2}{\line(1,0){150}} 

\multiput(0,0)(200,0){2}{\circle*{20}} 
\put(300,0){\circle*{20}} 
\multiput(150,86.6)(200,0){2}{\circle*{20}} 
\put(50,86.6){\circle*{20}} 
\multiput(0,173.2)(200,0){2}{\circle*{20}}
\put(300,173.2){\circle*{20}} 
\multiput(150,259.8)(200,0){2}{\circle*{20}}
\put(50,259.8){\circle*{20}} 
\multiput(0,346.4)(200,0){2}{\circle*{20}}
\put(300,346.4){\circle*{20}} 

\multiput(100,0)(0,173.2){3}{\circle{14}} 
\multiput(250,86.6)(0,173.2){2}{\circle{14}} 

\newsavebox{\diaguptri}
\savebox{\diaguptri}(100,100)[bl]{\begin{picture}(100,100)
\put(3.5,6.062){\line(3,5){44}}
\end{picture}}

\newsavebox{\diagdowntri}
\savebox{\diagdowntri}(100,100)[bl]{\begin{picture}(100,100)
\put(3.5,-6.062){\line(3,-5){44}}
\end{picture}}

\multiput(0,173.2)(50,86.6){2}{\usebox{\diaguptri}}
\multiput(0,0)(50,86.6){4}{\usebox{\diaguptri}}
\multiput(100,0)(50,86.6){4}{\usebox{\diaguptri}}
\multiput(200,0)(50,86.6){3}{\usebox{\diaguptri}}
\put(300,0){\usebox{\diaguptri}} 

\multiput(0,173.2)(50,-86.6){2}{\usebox{\diagdowntri}} 
\multiput(0,346.4)(50,-86.6){4}{\usebox{\diagdowntri}} 
\multiput(100,346.4)(50,-86.6){4}{\usebox{\diagdowntri}} 
\multiput(200,346.4)(50,-86.6){3}{\usebox{\diagdowntri}} 
\put(300,346.4){\usebox{\diagdowntri}} 

\thicklines

\newsavebox{\hortrisingle}
\savebox{\hortrisingle}(100,100)[bl]{\begin{picture}(200,100)
\put(0,0){\line(1,0){100}}
\end{picture}} 

\newsavebox{\diaguptrithick}
\savebox{\diaguptrithick}(100,100)[bl]{\begin{picture}(100,100)
\put(3.5,6.062){\line(3,5){44}}
\end{picture}}

\newsavebox{\diagdowntrithick}                         
\savebox{\diagdowntrithick}(100,100)[bl]{\begin{picture}(100,100)
\put(3.5,-6.062){\line(3,-5){44}}
\end{picture}}

\multiput(200,0)(0,173.2){3}{\usebox{\hortrisingle}} 
\multiput(50,86.6)(0,173.2){2}{\usebox{\hortrisingle}} 
\multiput(0,173.2)(150,-86.6){3}{\usebox{\diaguptrithick}} 
\multiput(150,259.8)(150,-86.6){2}{\usebox{\diaguptrithick}} 
\put(0,0){\usebox{\diaguptrithick}} 
\multiput(0,346.4)(150,-86.6){3}{\usebox{\diagdowntrithick}} 
\multiput(0,173.2)(150,-86.6){2}{\usebox{\diagdowntrithick}} 
\put(300,346.4){\usebox{\diagdowntrithick}} 

\put(360,260){\large $t_1^\prime$} 
\put(300,366){\large $t_2^\prime$} 
\put(200,366){\large $t_3^\prime$} 
\put(45,280){\large $t_4^\prime$} 
\put(-20,173){\large $t_5^\prime$} 
\put(45,66){\large $t_6^\prime$} 
\put(200,-20){\large $t_7^\prime$} 
\put(300,-20){\large $t_8^\prime$} 
\put(360,86.6){\large $t_9^\prime$} 

\put(210,180){\large $t_0$} 
\put(310,178){\large $t_1$} 
\put(145,280){\large $t_2$} 
\put(145,66){\large $t_3$} 

\end{picture}
\vspace*{1cm}
\caption{Decimation for the triangular lattice. The original triangular lattice
corresponds to both thin and thick lines;  
the empty circles represent the spins summed over; the hexagonal lattice 
resulting from decimation is drawn with solid circles and thick lines.  
Each spin $t_0$ of the decimated lattice 
interacts with 12 spins: three nearest neighbors 
$t_1,t_2,t_3$ and the nine next-to-nearest neighbors   
$t_1^\prime,\ldots,t_9^\prime$.  
} 
\label{figure_triangular_lattice}
\end{figure}  

\newpage 

%
%
%

\begin{figure}
\centering 
\begin{picture}(400,350)(0,0)  

\thicklines

\multiput(0,0)(0,173.2){3}{\line(1,0){400}} 
\multiput(50,0)(100,0){4}{\circle*{20}} 
\multiput(50,173.2)(100,0){4}{\circle*{20}} 
\multiput(50,346.4)(100,0){4}{\circle*{20}} 
\multiput(50,0)(100,0){4}{\line(0,1){346.4}}  
\multiput(0,86.6)(200,0){3}{\circle{14}} 
\multiput(100,259.8)(200,0){2}{\circle{14}} 

\thinlines 

\newsavebox{\diagupkag}
\savebox{\diagupkag}(100,100)[bl]{\begin{picture}(100,100)
\put(3.5,6.062){\line(3,5){44}}
\end{picture}}

\newsavebox{\diagdownkag}
\savebox{\diagdownkag}(100,100)[bl]{\begin{picture}(100,100)
\put(3.5,-6.062){\line(3,-5){44}}
\end{picture}} 

\multiput(0,86.6)(50,86.6){3}{\usebox{\diagupkag}} 
\multiput(150,0)(50,86.6){4}{\usebox{\diagupkag}}  
\put(350,0){\usebox{\diagupkag}}

\put(0,86.6){\usebox{\diagdownkag}}
\multiput(50,346.4)(50,-86.6){4}{\usebox{\diagdownkag}}  
\multiput(250,346.4)(50,-86.6){3}{\usebox{\diagdownkag}}  %

\put(260,153){\large $t_0$} 

\put(360,183){\large $t_1$} 
\put(360,325){\large $t_1^\prime$} 
\put(230,325){\large $t_2^\prime$} 
\put(130,153){\large $t_2$} 
\put(150,-20){\large $t_3^\prime$} 
\put(250,-20){\large $t_4^\prime$} 

\end{picture}
\vspace*{1cm}
\caption{Decimation for the Kagom\'e lattice. The original Kagom\'e lattice
corresponds to all the thin lines and the horizontal thick lines; 
the empty circles represent the spins summed over; the  
square lattice resulting from decimation is defined by the thick lines. 
Each spin $t_0$ of the decimated lattice (solid circles)    
interacts with 6 spins: two nearest neighbors $t_1, t_2$ and four 
next-to-nearest neighbors 
$t_1^\prime,t_2^\prime,t_3^\prime,t_4^\prime$. All these six spins 
live on the two ``crossed'' squares to which $t_0$ belongs.  
} 
\label{figure_kagome_lattice}
\end{figure}
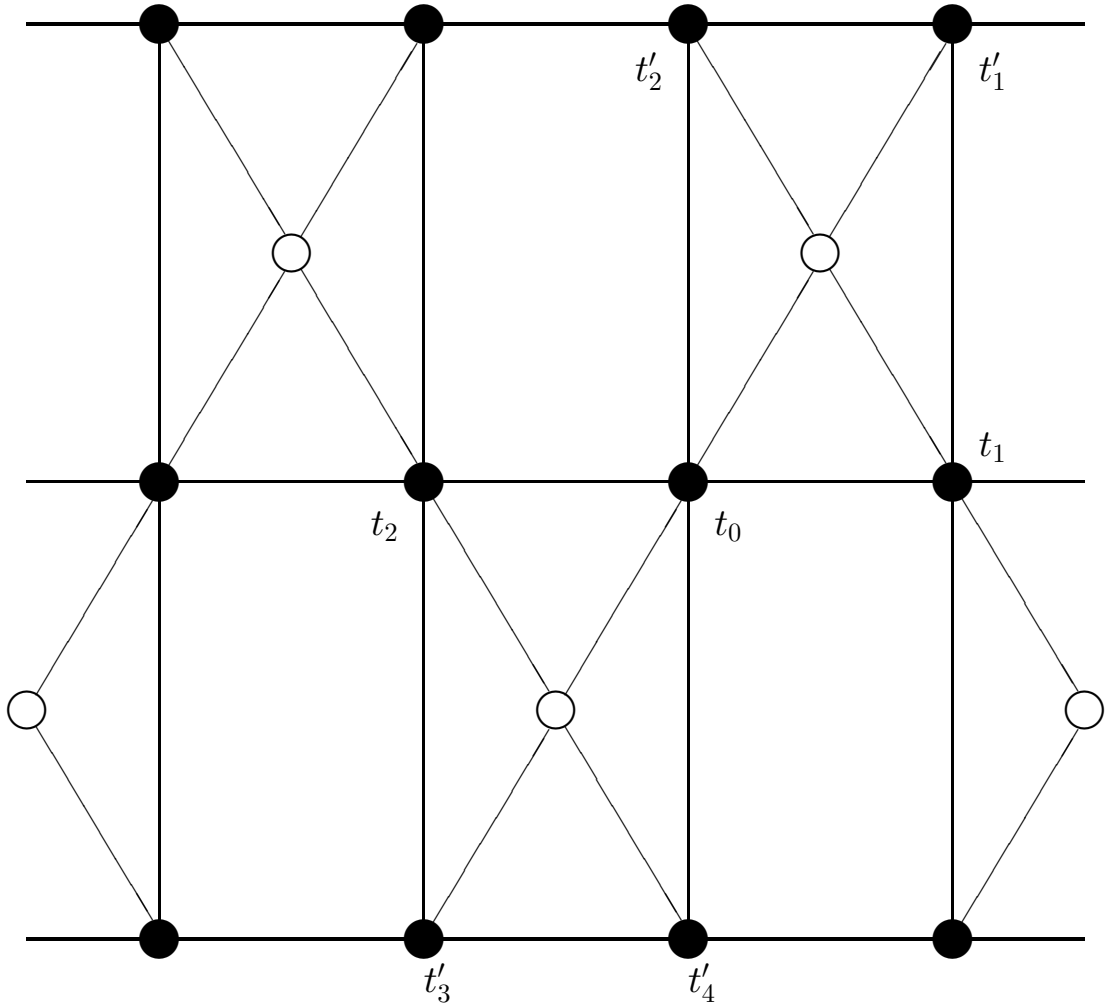  

\newpage 

%
%
%

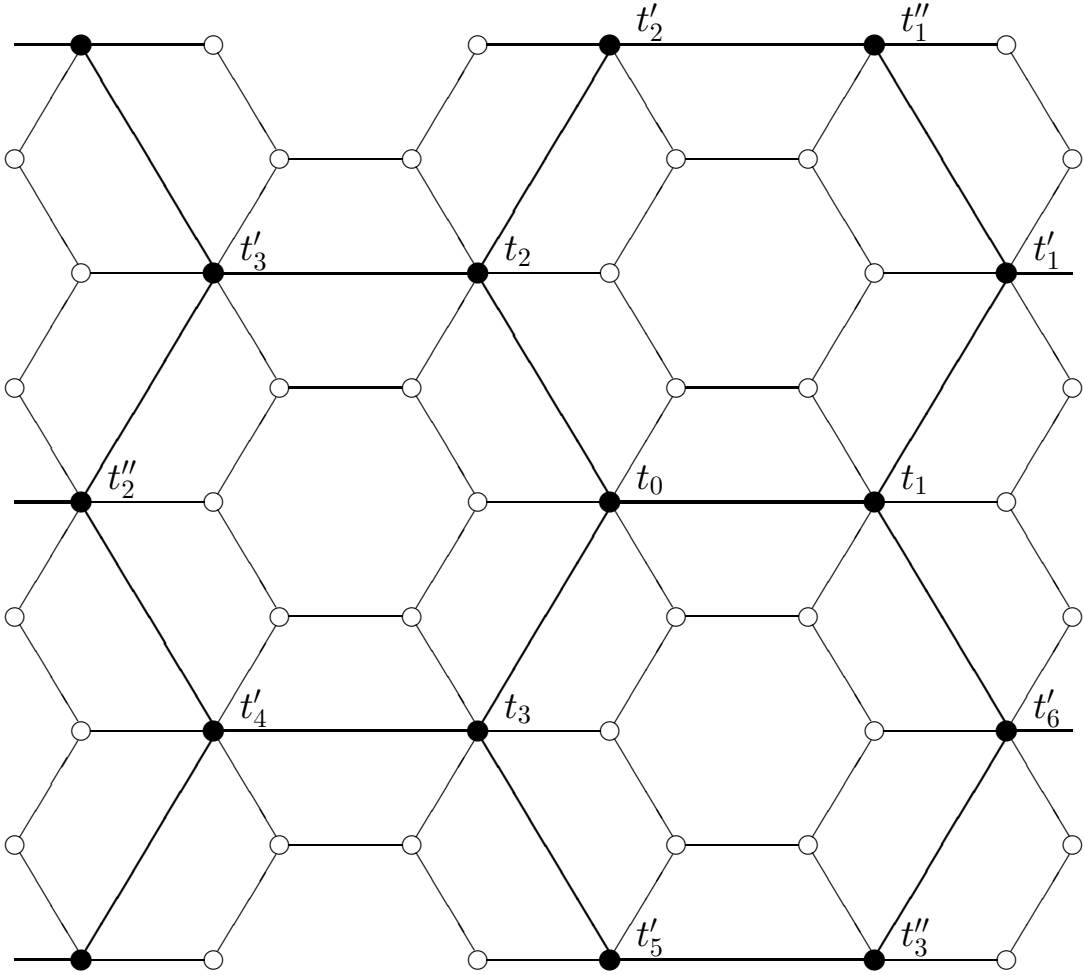
\begin{figure}
\setlength{\unitlength}{0.5pt} 
\centering 
\begin{picture}(800,700)(0,0) 

\thinlines
\newsavebox{\horhexbis}
\savebox{\horhexbis}(100,100)[bl]{\begin{picture}(100,100)
\put(7,0){\line(1,0){86}} 
\end{picture}}   

\multiput(50,0)(300,0){3}{\usebox{\horhexbis}} 
\multiput(50,173.2)(300,0){3}{\usebox{\horhexbis}} 
\multiput(50,346.4)(300,0){3}{\usebox{\horhexbis}} 
\multiput(50,519.6)(300,0){3}{\usebox{\horhexbis}} 
\multiput(50,692.8)(300,0){3}{\usebox{\horhexbis}} 

\multiput(200,86.6)(300,0){2}{\usebox{\horhexbis}}
\multiput(200,259.8)(300,0){2}{\usebox{\horhexbis}}
\multiput(200,433.0)(300,0){2}{\usebox{\horhexbis}}
\multiput(200,606.2)(300,0){2}{\usebox{\horhexbis}}

\multiput(0,86.5)(0,173.2){4}{\circle{14}} 
\multiput(200,86.5)(0,173.2){4}{\circle{14}} 
\multiput(300,86.5)(0,173.2){4}{\circle{14}} 
\multiput(500,86.5)(0,173.2){4}{\circle{14}} 
\multiput(600,86.5)(0,173.2){4}{\circle{14}} 
\multiput(800,86.5)(0,173.2){4}{\circle{14}} 

\multiput(50,0)(0,173.2){5}{\circle{14}} 
\multiput(150,0)(0,173.2){5}{\circle{14}} 
\multiput(350,0)(0,173.2){5}{\circle{14}} 
\multiput(450,0)(0,173.2){5}{\circle{14}} 
\multiput(650,0)(0,173.2){5}{\circle{14}} 
\multiput(750,0)(0,173.2){5}{\circle{14}} 

\multiput(50,0)(0,346.4){3}{\circle*{15}}
\multiput(450,0)(0,346.4){3}{\circle*{15}}
\multiput(650,0)(0,346.4){3}{\circle*{15}}
\multiput(150,173.2)(0,346.4){2}{\circle*{15}}
\multiput(350,173.2)(0,346.4){2}{\circle*{15}}
\multiput(750,173.2)(0,346.4){2}{\circle*{15}}

\newsavebox{\diagupbis}
\savebox{\diagupbis}(100,100)[bl]{\begin{picture}(100,100)
\put(3.5,6.062){\line(3,5){44}}
\end{picture}} 

\multiput(150,0)(-150,86.6){2}{\usebox{\diagupbis}}
\multiput(450,0)(-150,86.6){4}{\usebox{\diagupbis}}
\multiput(750,0)(-150,86.6){6}{\usebox{\diagupbis}}
\multiput(750,173.2)(-150,86.6){6}{\usebox{\diagupbis}}
\multiput(750,346.4)(-150,86.6){4}{\usebox{\diagupbis}}
\multiput(750,519.6)(-150,86.6){2}{\usebox{\diagupbis}}

\newsavebox{\diagdownbis}
\savebox{\diagdownbis}(100,100)[bl]{\begin{picture}(100,100)
\put(3.5,-6.062){\line(3,-5){44}}
\end{picture}} 

\multiput(0,86.6)(150,86.6){6}{\usebox{\diagdownbis}} 
\multiput(300,86.6)(150,86.6){4}{\usebox{\diagdownbis}} 
\multiput(600,86.6)(150,86.6){2}{\usebox{\diagdownbis}} 
\multiput(0,259.8)(150,86.6){6}{\usebox{\diagdownbis}} 
\multiput(0,433.0)(150,86.6){4}{\usebox{\diagdownbis}} 
\multiput(0,606.2)(150,86.6){2}{\usebox{\diagdownbis}} 

\thicklines 
\newsavebox{\horhexbiss}
\savebox{\horhexbiss}(100,100)[bl]{\begin{picture}(100,200)
\put(7,0){\line(1,0){186}}
\end{picture}}
\multiput(150,173.2)(0,346.4){2}{\usebox{\horhexbiss}} 
\multiput(450,0)(0,346.4){3}{\usebox{\horhexbiss}} 

\newsavebox{\diagupbiss}
\savebox{\diagupbiss}(100,100)[bl]{\begin{picture}(100,100)
\put(3.5,7){\line(3,5){96}}
\end{picture}} 
\put(50,0){\usebox{\diagupbiss}} 
\multiput(50,346.4)(300,-173.2){3}{\usebox{\diagupbiss}} 
\multiput(350,519.6)(300,-173.2){2}{\usebox{\diagupbiss}} 
 
\newsavebox{\diagdownbiss}
\savebox{\diagdownbiss}(100,100)[bl]{\begin{picture}(100,100)
\put(3.5,-7){\line(3,-5){96}}
\end{picture}} 
\put(50,692.8){\usebox{\diagdownbiss}} 
\multiput(50,346.4)(300,173.2){3}{\usebox{\diagdownbiss}}  
\multiput(350,173.2)(300,173.2){2}{\usebox{\diagdownbiss}} 

\put(0,0){\line(1,0){50}} 
\put(0,346.4){\line(1,0){50}} 
\put(0,692.8){\line(1,0){50}} 
\put(750,173.2){\line(1,0){50}} 
\put(750,519.6){\line(1,0){50}} 

\put(470,356.4){\large $t_0$} 
\put(670,356.4){\large $t_1$} 
\put(370,529.6){\large $t_2$} 
\put(370,183.2){\large $t_3$} 
\put(770,529.6){\large $t_1^\prime$}
\put(470,702.8){\large $t_2^\prime$}
\put(170,529.6){\large $t_3^\prime$}
\put(170,183.2){\large $t_4^\prime$}
\put(470,10){\large $t_5^\prime$}
\put(770,183.2){\large $t_6^\prime$}
\put(670,702.8){\large $t_1^{\prime\prime}$}
\put(70,356.4){\large $t_2^{\prime\prime}$}
\put(670,10){\large $t_3^{\prime\prime}$}

\end{picture}
\vspace*{1cm}
\caption{ 
More complex decimation for the hexagonal lattice. The original lattice is 
drawn with thin lines; the empty circles represent the spins summed over;  
the decimated lattice is drawn with solid circles 
and thick lines. Each spin $t_0$ of the decimated lattice interacts 
with 12 spins: three nearest neighbors $t_1,t_2,t_3$, six 
second-nearest neighbors $t_1^\prime,\ldots,t_6^\prime$, and 
three third-nearest neighbors    
$t_1^{\prime\prime},t_2^{\prime\prime},t_3^{\prime\prime}$.   
} 
\label{figure_hexagonal_lattice_bis}
\end{figure}  

\newpage 

%
%
%

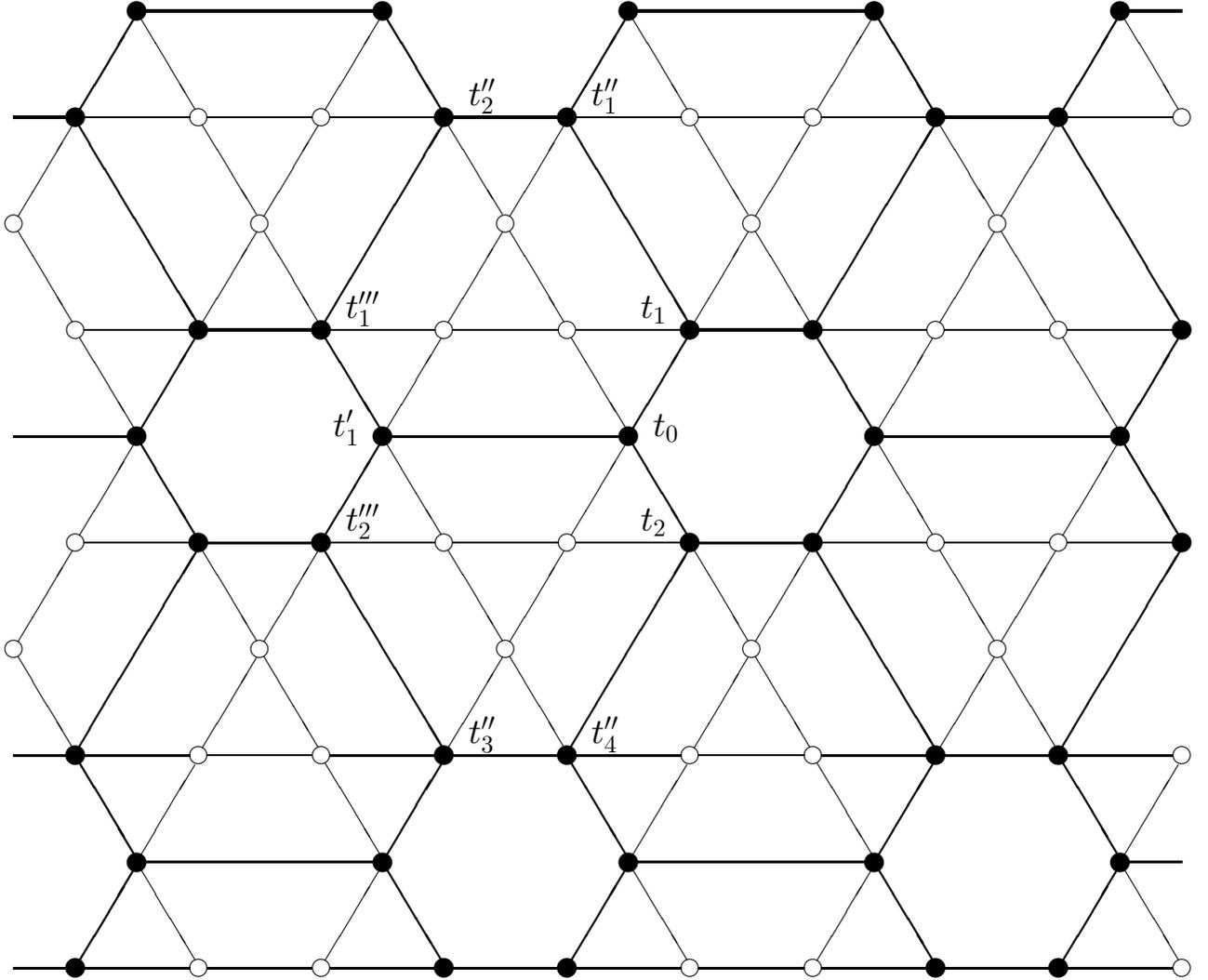
\begin{figure}
\setlength{\unitlength}{0.5pt} 
\centering 
\begin{picture}(950,800)(0,0)  

\thicklines  
\multiput(100,86.5)(400,0){2}{\line(1,0){200}} 
\multiput(100,779.4)(400,0){2}{\line(1,0){200}} 
\multiput(300,433.0)(400,0){2}{\line(1,0){200}} 

\newsavebox{\hexthick}
\savebox{\hexthick}(200,200)[bl]{\begin{picture}(200,200)
\put(3.5,92.562){\line(3,5){44}}
\put(153.5,6.062){\line(3,5){44}}
\put(3.5,80.438){\line(3,-5){44}}
\put(153.5,167.138){\line(3,-5){44}}
\put(57,0){\line(1,0){86}}
\put(57,173.2){\line(1,0){86}}
\put(50,0){\circle*{15}} 
\put(150,0){\circle*{15}} 
\put(0,86,5){\circle*{15}} 
\put(200,86,5){\circle*{15}} 
\put(50,173.2){\circle*{15}} 
\put(150,173.2){\circle*{15}} 
\end{picture}}

\thinlines 
\multiput(50,0)(100,0){10}{\circle{14}} 
\multiput(50,173.2)(100,0){10}{\circle{14}} 
\multiput(50,346.4)(100,0){10}{\circle{14}} 
\multiput(50,519.6)(100,0){10}{\circle{14}} 
\multiput(50,692.8)(100,0){10}{\circle{14}} 

\multiput(100,779.4)(200,0){5}{\circle*{15}} 
\put(900,433.0){\circle*{15}} 
\put(100,86.5){\circle*{15}} 
\multiput(0,259.8)(200,0){5}{\circle{14}} 
\multiput(0,606.2)(200,0){5}{\circle{14}} 

\newsavebox{\hexthin}
\savebox{\hexthin}(200,200)[bl]{\begin{picture}(200,200)
\put(3.5,92.562){\line(3,5){44}}
\put(153.5,6.062){\line(3,5){44}}
\put(3.5,80.438){\line(3,-5){44}}
\put(153.5,167.138){\line(3,-5){44}}
\put(57,0){\line(1,0){86}} 
\put(57,173.2){\line(1,0){86}} 
\end{picture}}

\multiput(300,0)(400,0){2}{\usebox{\hexthick}} 
\multiput(100,346.4)(400,0){2}{\usebox{\hexthick}} 
\multiput(100,0)(400,0){2}{\usebox{\hexthin}} 
\multiput(0,173.2)(200,0){4}{\usebox{\hexthin}} 
\multiput(300,346.4)(400,0){2}{\usebox{\hexthin}} 
\multiput(0,519.6)(200,0){4}{\usebox{\hexthin}} 

\newsavebox{\horkagbis}
\savebox{\horkagbis}(100,100)[bl]{\begin{picture}(100,100)
\put(7,0){\line(1,0){86}}
\end{picture}} 

\multiput(50,0)(200,0){5}{\usebox{\horkagbis}} 
\multiput(850,173.2)(0,173.2){4}{\usebox{\horkagbis}} 
\multiput(150,692.8)(200,0){4}{\usebox{\horkagbis}}

\newsavebox{\diagupkagbis}
\savebox{\diagupkagbis}(100,100)[bl]{\begin{picture}(100,100)
\put(3.5,6.062){\line(3,5){44}}
\end{picture}}

\multiput(50,0)(0,346.4){3}{\usebox{\diagupkagbis}} 
\multiput(900,86.5)(0,346.4){2}{\usebox{\diagupkagbis}} 
\multiput(250,692.8)(200,0){4}{\usebox{\diagupkagbis}} 
\multiput(800,259.8)(0,346.4){2}{\usebox{\diagupkagbis}} 

\newsavebox{\diagdownkagbis}
\savebox{\diagdownkagbis}(100,100)[bl]{\begin{picture}(100,100)
\put(3.5,-6.062){\line(3,-5){44}}
\end{picture}} 

\multiput(50,173.2)(0,346.4){2}{\usebox{\diagdownkagbis}} 
\multiput(800,259.8)(0,346.4){2}{\usebox{\diagdownkagbis}} 
\multiput(900,86.5)(0,346.4){3}{\usebox{\diagdownkagbis}} 
\multiput(100,779.4)(200,0){4}{\usebox{\diagdownkagbis}} 

\thicklines 

\newsavebox{\diagupkagbiss}
\savebox{\diagupkagbiss}(100,100)[bl]{\begin{picture}(100,100)
\put(3.5,6.062){\line(3,5){44}}
\end{picture}} 

\newsavebox{\diagdownkagbiss}
\savebox{\diagdownkagbiss}(100,100)[bl]{\begin{picture}(100,100)
\put(3.5,-6.062){\line(3,-5){44}}
\end{picture}} 

\newsavebox{\horkagbiss}
\savebox{\horkagbiss}(100,100)[bl]{\begin{picture}(100,100)
\put(7,0){\line(1,0){86}}
\end{picture}}

\multiput(50,692.8)(400,0){3}{\usebox{\diagupkagbiss}} 
\multiput(300,779.4)(400,0){2}{\usebox{\diagdownkagbiss}} 
\put(900,433.0){\usebox{\diagupkagbiss}} 
\put(900,433.0){\usebox{\diagdownkagbiss}} 
\put(50,0){\usebox{\diagupkagbiss}} 
\put(50,173.2){\usebox{\diagdownkagbiss}} 
\put(350,692.8){\usebox{\horkagbiss}} 
\put(750,692.8){\usebox{\horkagbiss}} 

\multiput(50,692.8)(400,0){3}{\circle*{15}} 
\multiput(350,692.8)(400,0){2}{\circle*{15}} 
\put(950,519.6){\circle*{15}} 
\put(950,346.4){\circle*{15}} 
\put(50,0){\circle*{15}} 
\put(50,173.2){\circle*{15}} 

\newsavebox{\diagupkagtris}     
\savebox{\diagupkagtris}(100,100)[bl]{\begin{picture}(100,100)
\put(0,2){\line(3,5){100}}
\end{picture}}

\newsavebox{\diagdownkagtris}
\savebox{\diagdownkagtris}(100,100)[bl]{\begin{picture}(100,100)
\put(0,-2){\line(3,-5){100}}
\end{picture}}

\put(50,173.2){\usebox{\diagupkagtris}} 
\put(250,519.6){\usebox{\diagupkagtris}} 
\put(450,173.2){\usebox{\diagupkagtris}} 
\put(650,519.6){\usebox{\diagupkagtris}} 
\put(850,173.2){\usebox{\diagupkagtris}} 

\put(50,692.8){\usebox{\diagdownkagtris}} 
\put(250,346.4){\usebox{\diagdownkagtris}} 
\put(450,692.8){\usebox{\diagdownkagtris}} 
\put(650,346.4){\usebox{\diagdownkagtris}} 
\put(850,692.8){\usebox{\diagdownkagtris}} 

\put(0,0){\line(1,0){50}} 
\put(0,173.2){\line(1,0){50}} 
\put(0,433.0){\line(1,0){100}} 
\put(0,692.8){\line(1,0){50}} 

\put(900,86.5){\line(1,0){50}} 
\put(900,779.4){\line(1,0){50}} 

\put(520,433.0){\large $t_0$} 
\put(510,529.6){\large $t_1$} 
\put(510,356.4){\large $t_2$} 
\put(260,433.0){\large $t_1^\prime$} 
\put(470,702.8){\large $t_1^{\prime\prime}$} 
\put(370,702.8){\large $t_2^{\prime\prime}$} 
\put(370,183.2){\large $t_3^{\prime\prime}$} 
\put(470,183.2){\large $t_4^{\prime\prime}$} 
\put(270,529.6){\large $t_1^{\prime\prime\prime}$} 
\put(270,356.4){\large $t_2^{\prime\prime\prime}$}

\end{picture}
\vspace*{1cm}
\caption{  
More complex decimation for the Kagom\'e lattice. The original lattice 
corresponds to the whole set of circles (empty and solid) and the 
{\em nearest-neighbor} bonds between them; the empty circles 
represent the spins summed over;  
the decimated lattice is drawn with solid circles
and thick lines. Each spin $t_0$ of the decimated lattice interacts
with nine spins: two nearest neighbors $t_1,t_2$, one  
second-nearest neighbor $t_1^\prime$, four third-nearest neighbors  
$t_1^{\prime\prime},t_2^{\prime\prime},t_3^{\prime\prime},t_4^{\prime\prime}$,
and two fourth-nearest neighbors  
$t_1^{\prime\prime\prime},t_2^{\prime\prime\prime}$. 
} 
\label{figure_kagome_lattice_bis}
\end{figure}



\addcontentsline{toc}{section}{References}
\begin{thebibliography}{199}

\bibitem{Dobrushin_68} R.L. Dobrushin, Theor. Prob. Appl. {\bf 13}, 197 (1968).

\bibitem{Dobrushin_70} R.L. Dobrushin, Theor. Prob. Appl. {\bf 15}, 458 (1970).

\bibitem{Georgii_88} H.-O. Georgii, {\em Gibbs Measures and Phase Transitions}\/
       (de Gruyter, Berlin--New York, 1988).

\bibitem{Simon_93} B. Simon, {\em The Statistical Mechanics of Lattice Gases}\/
       (Princeton University Press, Princeton, NJ, 1993).

\bibitem{Potts_52}  R.B. Potts, Proc. Camb. Philos. Soc. {\bf 48}, 106 (1952).

\bibitem{Wu_82}  F.Y. Wu, Rev. Mod. Phys. {\bf 54}, 235 (1982); {\bf 55},
  315 (E) (1983).

\bibitem{Wu_84}  F.Y. Wu, J. Appl. Phys. {\bf 55}, 2421 (1984).

\bibitem{Stephenson_64}  J. Stephenson, J. Math. Phys. {\bf 5}, 1009 (1964). 

\bibitem{Syozi_72} I. Syozi, in
   {\em Phase Transition and Critical Phenomena}\/, Vol. 1,
   edited by C.~Domb and M.S.~Green (Academic Press, London, 1972). 


\bibitem{Baxter_82}  R.J. Baxter, {\em Exactly Solved Models in Statistical
   Mechanics}\/ (Academic Press, London--New York, 1982).

\bibitem{Baxter_82b} R.J. Baxter, Proc. Roy. Soc. London {\bf A383}, 43 (1982).

\bibitem{Onsager_44}  L. Onsager, Phys. Rev. {\bf 65}, 117 (1944).

\bibitem{Lenard_67}  A. Lenard, cited in E.H. Lieb, Phys. Rev. {\bf 162}, 162
   (1967) at pp.~169--170.

\bibitem{Baxter_70}  R.J. Baxter, J. Math. Phys. {\bf 11}, 3116 (1970).

\bibitem{Truong_86}  T.T. Truong and K.D. Schotte, J. Phys. {\bf A19}, 1477
   (1986).

\bibitem{Pearce_89a}  P.A. Pearce and K.A. Seaton, Ann. Phys. {\bf 193}, 326
   (1989).

\bibitem{Pearce_89b}  D. Kim and P.A. Pearce, J. Phys. {\bf A22}, 1439 (1989).

\bibitem{Henley_94}  C.L. Henley, in preparation;
   J.K. Burton and C.L. Henley, in preparation.

\bibitem{Ferreira-Sokal}  S.J. Ferreira and A.D. Sokal, Phys. Rev. {\bf B51},
   6727 (1995);  and in preparation.

\bibitem{Saleur_90}  H. Saleur, Commun. Math. Phys. {\bf 132}, 657 (1990).

\bibitem{Saleur_91}  H. Saleur, Nucl. Phys. {\bf B360}, 219 (1991).


\bibitem{Baxter_78}  R.J. Baxter, H.N.V. Temperley and S.E. Ashley,
   Proc. Roy. Soc. London {\bf A358}, 535 (1978).

\bibitem{Baxter_86}  R.J. Baxter, J. Phys. {\bf A19}, 2821 (1986).
 
\bibitem{Baxter_87}  R.J. Baxter, J. Phys. {\bf A20}, 5241 (1987).

\bibitem{Baxter_70_TRI}  R.J. Baxter, J. Math. Phys. {\bf 11}, 784 (1970).

\bibitem{vEFS_unpub} A. van Enter, R. Fern\'andez and A.D. Sokal, unpublished. 

\bibitem{Adler_95}  J. Adler, A. Brandt, W. Janke and S. Shmulyan,
   J. Phys. {\bf A28}, 5117 (1995).

\bibitem{Fortuin-Kasteleyn_69} P.W. Kasteleyn and C.M. Fortuin, J. Phys. Soc.
Japan {\bf 26} (Suppl.), 11 (1969).  

\bibitem{Fortuin-Kasteleyn_72} C.M. Fortuin and P.W. Kasteleyn, 
Physica {\bf 57}, 536 (1972).   

\bibitem{Fortuin_72} C.M. Fortuin, Physica {\bf 58}, 393 (1972); 
 {\bf 59}, 545 (1972).  

\bibitem{Baxter_76}  R.J. Baxter, S.B. Kelland and F.Y. Wu,
   J. Phys. {\bf A9}, 397 (1976). 


\bibitem{Takano_71} M. Takano, T. Shinjo and T. Takada, J. Phys. Soc. Japan 
{\bf 30}, 1049 (1971). 

\bibitem{Broholm_91}  C. Broholm, G. Aeppli, G. Espinosa and A.S. Cooper,
    J. Appl. Phys. {\bf 69}, 4968 (1991).

\bibitem{Huse_92}  D. Huse and A.D. Rutemberg, Phys. Rev. {\bf B45},
   7536 (1992).

\bibitem{Henley_95} J. Kondev and C.L. Henley, ``Kac--Moody Symmetries
of Critical Ground States'', {\tt cond-mat/9511102}. 

%
%
%
%
%
%

\bibitem{vEFS_93} A.C.D. van Enter, R. F\'ernandez and A.D. Sokal, 
J.~Stat. Phys. {\bf 72}, 879 (1993).  

\bibitem{Preston_76} C. Preston, {\em Random Fields} (Springer-Verlag,
Berlin, 1976). 

\bibitem{Kennedy_95} K. Haller and T. Kennedy, ``Absence of Renormalization
Group Pathologies near the Critical Temperature -- Two Examples'',
University of Arizona
preprint, {\tt mp-arc/95-505}. 

\end{thebibliography}
\end{document}